\documentclass[aps,prb, amsmath,amssymb,floatfix,12pt, longbibliography]{revtex4-1}
\usepackage{tabularx}
\usepackage{bm}
\usepackage{euscript}
\usepackage{graphicx}
\usepackage{color}
\usepackage{amsfonts}
\usepackage{exscale}
\usepackage{amsbsy}
\usepackage{subfigure}
\usepackage{textcomp}
\usepackage{comment}
\usepackage{refcount}

\usepackage{hyperref}

\numberwithin{equation}{section}

\newcommand{\sgn}{\mathrm{sgn}}

\newcommand{\be}{\begin{equation}}
\newcommand{\ee}{\end{equation}}
\newcommand{\bea}{\begin{eqnarray}}
\newcommand{\eea}{\end{eqnarray}}
\newcommand{\tr}{\textrm{tr}}

\begin{document}

\title{Non-Fermi liquids at finite temperature: \\ normal state and infrared singularities}
\author{Huajia Wang$^{\Delta}$, Gonzalo Torroba$^{\phi}$}
\affiliation{$^\Delta$ Department of Physics, University of Illinois, Urbana IL, USA}
\affiliation{$^\phi$Centro At\'omico Bariloche and CONICET, Bariloche, Rio Negro R8402AGP,
Argentina}
\date{\today}

\begin{abstract}
We analyze quantum criticality at finite temperature for a class of non-Fermi liquids with massless bosons. Finite temperature gives rise to new infrared singularities that invalidate standard perturbative treatments. We show how such divergences are resolved at a nonperturbative level, and obtain the resulting fermion self-energy. This leads to a new ``thermal'' non-Fermi liquid regime that extends over a wide range of frequencies, and which violates finite temperature scaling laws near quantum critical points. We analyze the resulting quantum critical region and properties of the retarded Green's function. More generally, such effects dominate in the nearly static limit and are expected to have a nontrivial impact on superconductivity and transport.
\end{abstract}

\maketitle

\tableofcontents

%%%%%%%%%%%%%%%%%%%%%%%%%
%%%%%%%%%%%%%%%%%%%%%%%%%
%%%%%%%%%%%%%%%%%%%%%%%%%
%%%%%%%%%%%%%%%%%%%%%%%%%
\section{Introduction}\label{sec:intro}

\begin{figure}[h!]
\centering
\includegraphics[width=0.5\textwidth]{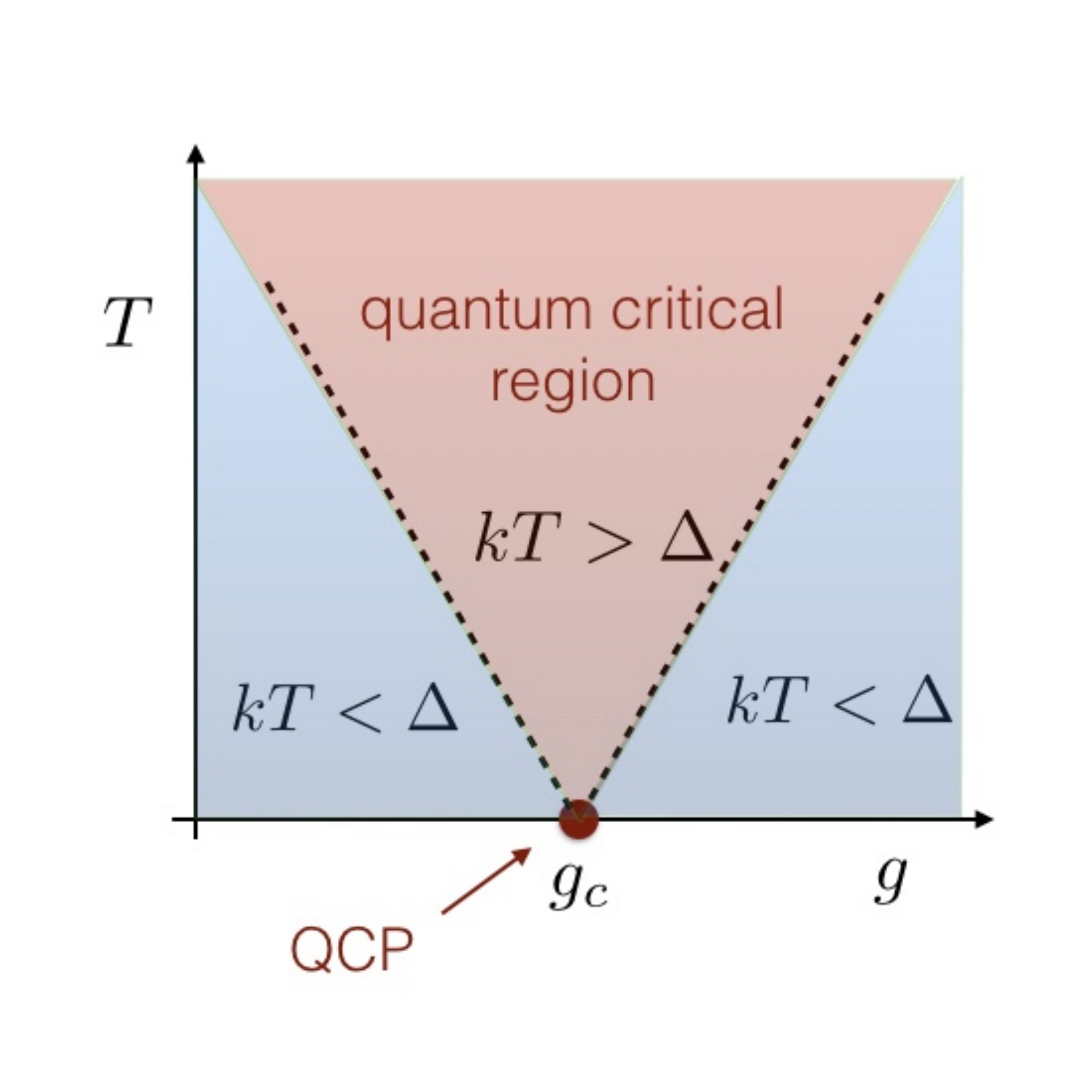}
\caption{Schematic phase diagram around a quantum critical point for coupling $g=g_c$. The quantum critical region (QCR) occurs for $kT>\Delta$, with $\Delta$ the typical mass scale for $g \neq g_c$.}\label{fig:qcr}
\end{figure}

There are strong indications that quantum critical points (QCP) are one of the organizing principles behind the behavior of strongly correlated materials \cite{Stewart2001, Broun2008, Gegenwart2008, taillefer-review, Shibauchi2013}. Although the QCP at zero temperature may be inaccessible experimentally, its influence extends into a whole quantum critical regime at finite $T$; this is illustrated in Fig. \ref{fig:qcr}. In this region, quantum and thermal effects compete, and can lead to dynamics strikingly different from that of weakly interacting quasiparticles. This could explain the origin of the ``strange metal'' regime of metallic systems.

The quantum critical region (denoted in what follows as QCR) should admit a field theoretic description, since the characteristic correlation length is much larger than the inverse temperature. One of the central problems of condensed matter physics is then to develop a field theory framework to explain the non-Fermi liquid dynamics in the QCR. This has been hampered by the fact that, in most models, superconductivity dominates over non-Fermi liquid effects; see \cite{Metlitski} for a recent discussion. When this happens, the physics above the superconducting dome is that of weakly coupled quasiparticles, and this is not useful for describing non-Fermi liquid properties of many metallic systems in their normal state.

The goal of this work is to study non-Fermi liquids (NFLs) at finite temperature. We consider systems near symmetry breaking transitions, where fluctuations of the order parameter give rise to relevant interactions on the Fermi surface. In order to address the issue of superconducting instabilities raised above, we focus on the large $N$ limit developed in \cite{Fitzpatrickone, FKKRtwo, Torroba:2014gqa, Raghu:2015sna}, where the fermion $\psi_i$ is promoted to a vector of $N$ components and the scalar $\phi_{ij}$ is an $N \times N$ matrix. We consider this large $N$ limit because it gives rise to a QCP that is stable to superconductivity at zero temperature\cite{Wang:2016hir}. Therefore, this parameter $N$ will allow us to access the QCR in a controlled way, and gives a new theoretical window into strange metallic behavior. We will focus on the behavior of quantum critical metals at finite temperature in their normal state. In a forthcoming work we consider 4-fermion interactions and the fate of superconductivity.\cite{future}

As it turns out, non-Fermi liquids at finite temperature suffer from new infrared singularities that are not present at $T=0$. Therefore, the dynamics in the QCR with $T>0$ could in principle be quite different from that at the $T=0$ critical point. A simple way to understand this is to realize that the exchange of virtual light bosons, present near QCPs, leads to a low-frequency singularity from the Bose-Einstein distribution,
\be\label{eq:IRdiv}
n_B(\omega) = \frac{1}{e^{\omega/kT}-1} \sim \frac{T}{\omega}\,.
\ee
This invalidates standard perturbative approaches. This problem was recognized already in early works on QED and QCD at finite density (see e.g.\cite{le2000thermal} for a review); Ref.\cite{Blaizot:1996az} showed how to resolve them by resumming infinite series of diagrams within the Bloch-Nordsiek approximation. This problem was also addressed for antiferromagnetic models in\cite{abanov2003quantum}. The large $N$ limit that we employ here will be essential for resolving IR singularities, as it implies that only rainbow diagrams need to be taken into account.

After reviewing in \S \ref{sec:model} the relevant properties of NFLs at zero temperature, in \S \ref{sec:one-loop} we will first consider quantum effects at finite $T$ within a one loop approximation. We will discuss in detail the structure of the thermal fermionic Green's function and identify the origin of the infrared singularities. The main part of the work is presented in \S \ref{sec:nonpert}. Using the large $N$ limit, we obtain a closed set of finite temperature Schwinger-Dyson equations for the bosonic and fermionic propagators in the normal state. We first argue that bosonic self-energy is not modified beyond one loop, generalizing to finite $T$ the result of\cite{Polchinski1994}; this gives  an overdamped boson with $z=3$ dynamical exponent. Next we focus on the fermionic self-energy. The IR divergences are resolved by rainbow resummation, giving rise to a new contribution to the self-energy that is independent of frequency. This results into a new ``thermal non-Fermi liquid'' regime, which is absent from the $T=0$ QCP. It violates the scaling dictated by the QCP. The phase structure and crossover regimes of the quantum critical region are
presented in \S \ref{sec:implications}. We also address some of the implications of our analysis, e.g. on real time dynamics and superconductivity, and discuss future directions. The Appendix contains explicit calculations used in the main text.

%%%%%%%%%%%%%%%%%%%%%%%%%
%%%%%%%%%%%%%%%%%%%%%%%%%
%%%%%%%%%%%%%%%%%%%%%%%%%
%%%%%%%%%%%%%%%%%%%%%%%%%
\section{Non-Fermi liquids near symmetry breaking transitions}\label{sec:model}

Let us review the class of NFLs that will be studied in the rest of the work. These arise from metals near a symmetry breaking transition at zero momentum. Assuming for concreteness that the global symmetry group is $SU(N)$, we take a Fermi surface of fermions $\psi_i$ transforming in the fundamental, coupled to a nearly massless $N\times N$ scalar $\phi_i^j$ in the adjoint representation. We work at large $N$ and in an $\epsilon$ expansion near three spatial dimensions (the critical dimension)\cite{Fitzpatrickone, FKKRtwo, Torroba:2014gqa, Raghu:2015sna}. Note that this large $N$ limit differs from the more studied theories with a large number of fermionic flavors \cite{Polchinski1994, Altshuler1994, Lee2009, Metlitski}; these models (as opposed to the matrix large $N$ ones) do not admit QCPs stable under superconductivity and hence are not suited for our purpose.

The euclidean finite temperature lagrangian takes the form
\be\label{eq:L1}
L=\frac{1}{2}\tr \left(\frac{1}{c^2}(\partial_\tau \phi)^2+ (\vec 
\nabla \phi)^2 \right)+ \psi_i^\dag \left(\partial_\tau+ \varepsilon(i\vec \nabla)-\mu_F
\right)\psi^i + \frac{g}{\sqrt{N}} \phi^i_j \psi^\dag_i \psi^j\,.
\ee
Since our focus here will be on the normal state, we ignore the 4-Fermi BCS interaction; superconductivity within the QCR will be analyzed in a forthcoming work.

Let us discuss the different terms in this action. The boson speed $c$ depends on the class of models. For instance, quantum criticality in hadronic matter arises from relativistic bosons (the gluons) and $c=1$. For metallic systems near symmetry breaking transitions, the scalar is emergent and starts off with only the gradient term; hence $c \to \infty$. In any case, as we review below, the boson inherits its dynamics from fermionic quantum effects (the Landau damping function), and this dominates at low energies over a possible tree level term with time derivatives. Furthermore, since we want to analyze the QCR, we will always tune the physical boson mass to zero. A possible $\phi^4$ interaction is not important for our discussion, and is set to zero.

Consider next the fermion terms. We assume a smooth Fermi surface, and approximate it locally by a sphere of radius $k_F$ (the Fermi momentum). We focus on the low energy dynamics near the Fermi surface, and expand the fermionic momentum as
\be\label{eq:spherical}
\vec p = \hat n (k_F+p_\perp)\;,\;p_\perp \ll k_F
\ee
with $\hat n$ a unit vector that fixes the position on the Fermi surface, and $\varepsilon(k_F) = \mu_F$. See e.g. \cite{Shankar}. At linear order, $ \varepsilon(\vec p ) -\mu_F \approx v p_\perp $, where the Fermi velocity $v = \varepsilon'(k_F)$.
Finally, the most relevant boson-fermion interaction term compatible with the $SU(N)$ global symmetry is the Yukawa coupling in (\ref{eq:L1}). The theory admits a consistent large $N$ limit if $g$ is held fixed. 

Refs.\cite{Raghu:2015sna, Wang:2016hir} obtained the dynamics at $T=0$, taking into account both the cubic Yukawa term as well as the BCS coupling and potential superconducting instabilites. For $N\ll N_c=\frac{12}{\epsilon}$, the system is in a weakly coupled SC phase. Increasing $N$ towards $N_c$ enhances NFL effects, such as the fermionc anomalous dimension. Superconductivity is destroyed by incoherent fluctuations at $N= N_c$ through an infinite order transition, and for $N>N_c$ the system flows to a stable QCP. The large $N$ exact fermion Green's function at the fixed point reads
\be\label{eq:GFQCP}
G_F(\omega, p) =- \frac{1}{i \Lambda_{NFL}^{2\gamma} \omega^{1-2\gamma}-\varepsilon_p}
\ee
with anomalous dimension
\be
\gamma= \frac{\epsilon}{6}\,.
\ee
The scale $\Lambda_{NFL}$ determines the cross-over between the FL and NFL regimes. The Yukawa coupling also flows to a nontrivial value
\be
\alpha = \frac{g^2}{12\pi^2 v} \to \frac{\epsilon}{3}\,.
\ee
The 4-Fermi coupling $\lambda_{BCS}$ also flows to a fixed point --NFL effects are sufficiently strong to stop its flow towards $\lambda_{BCS} \to \infty$ (which would correspond to a superconducting instability).

Our task in what follows will be to analyze the NFL behavior at finite temperature. We will find that the $T=0$ QCP controls the high frequency regime $\omega \gg \pi T$. However, we will also uncover a new ``thermal'' NFL regime that arises for euclidean frequencies $\omega \sim \pi T$, or real frequencies $p_0 \ll T$.

%%%%%%%%%%%%%%%%%%%%%%%%%
%%%%%%%%%%%%%%%%%%%%%%%%%
%%%%%%%%%%%%%%%%%%%%%%%%%
%%%%%%%%%%%%%%%%%%%%%%%%%
\section{One loop analysis}\label{sec:one-loop}

We begin our analysis of the QCR by studying quantum effects at one loop. Later on we will extend this to all orders.

The dominant effects at large $N$ come from the boson and fermion self-energies, shown in Fig. \ref{fig:SDeq} (but with the exact propagators replaced by the tree-level ones). Corrections to the cubic interaction are suppressed by $1/N$, and will be neglected in what follows. Denoting the self-energies by $\Pi(\omega_n, p)$ and $\Sigma(\omega_n, p)$, their effect is to correct the bosonic and fermionic two-point functions as follows:
\bea\label{eq:fullGs}
D(\omega_n, p) &=& \frac{1}{\frac{1}{c^2}\omega_n^2 + p^2 +\Pi(\omega_n, p)}\nonumber\\
G(\omega_n, p) &=& -\frac{1}{i \omega_n- v p_\perp +i \Sigma(\omega_n, p)}\,.
\eea
Bosonic and fermionic Matsubara frequencies are given by $\omega_n= 2n \pi T$ and $\omega_n=(2n+1)\pi T$; we also work near the Fermi surface, linearizing the fermionic dispersion relation in terms of (\ref{eq:spherical}).

%%%%%%%%%%%%%%%%%%%%%%%%%%%%%%%%%%%%%%%
%%%%%%%%%%%%%%%%%%%%%%%%%%%%%%%%%%%%%%%
\subsection{Landau damping at finite temperature}\label{subsec:Pi1}

Although the evaluation of the Landau damping function is well known, it is useful to discuss it here to understand the energy scales involved, and for later use at higher loops. The loop integral for $d=3$ space dimensions, in the spherical decomposition (\ref{eq:spherical}) reads
\be
\Pi(\omega_n, q) = \frac{g^2 k_F^2}{(2\pi)^2 N}\, T \sum_m  \int_{-1}^1 d(\cos \theta)\,\int_{-\Lambda}^\Lambda dp_\perp\,\frac{1}{i\omega_m - v p_\perp}\frac{1}{i(\omega_m+\omega_n)-v (p_\perp +q \cos \theta)}\,,
\ee
with $\Lambda$ the momentum cutoff near the Fermi surface. Note that $\omega_n$ is bosonic while $\omega_m$ is fermionic.

Due to the UV divergence, the integral depends on the order of integration. Let us integrate first over $p_\perp$ in the limit $\Lambda \to \infty$, which gives a factor of $\frac{\Theta(\omega_m+\omega_n)-\Theta(\omega_m)}{i \omega_n- v q \cos \theta}$ (with $\Theta(x)$ the Heaviside function). The sum over $m$ and angular integral can then be performed explicitly, obtaining
\be\label{eq:Pi1}
\Pi(\omega_n, q) =\frac{2}{\pi} M_D^2\, \frac{\omega_n}{q}\,\tan^{-1}\left(\frac{v q}{\omega_n}\right)\,,
\ee
where we have introduced the Debye mass scale $M_D^2 \equiv \frac{g^2k_F^2}{4\pi v N}$. This agrees with the zero temperature result \cite{Torroba:2014gqa}, so thermal effects do not affect Landau damping at one loop. If instead the Matsubara sum is performed first (by rewriting it in terms of a complex integral), the final result differs from (\ref{eq:Pi1}) by a extra $-M_D^2$ term.\cite{nagaosa2013quantum} However, the important point is that this is again independent of temperature and can be set to zero by an appropriate choice of bare mass. Therefore, if we tune the boson to be close to the critical point at zero temperature, no thermal mass is generated at one loop, and no further tuning is needed.

While (\ref{eq:Pi1}) is formally a $1/N$ effect in our large $N$ expansion, it cannot be neglected because it is always relevant at low energies. In order to capture this, we take large $N$ but with $M_D$ fixed (more precisely, $\omega/M_D$ fixed). For $\omega < M_D$, the frequency dependence is dominated by the Landau damping term, and we obtain a boson with dynamical exponent $z=3$:
\be
D(\omega_n, q) \approx \frac{1}{q^2+ M_D^2\, \frac{|\omega_n|}{v q}}\,.
\ee
In what follows we work with this resumed boson propagator. We refer the reader to \cite{Torroba:2014gqa, Raghu:2015sna} for a more detailed discussion of this approach.

%%%%%%%%%%%%%%%%%%%%%%%%%%%%%%%%%%%%%%%
%%%%%%%%%%%%%%%%%%%%%%%%%%%%%%%%%%%%%%%
\subsection{Fermion self-energy}\label{subsec:Sigma1}

We now move on to discuss the much more subtle fermionic self-energy, with loop integral
\be\label{eq:Sigmaone0}
i \Sigma(\omega_n, p) = -g^2 T \sum_m \int \frac{d^d q}{(2\pi)^d}\,D(\omega_n-\omega_m, q-p)\,\frac{1}{i \omega_m-\varepsilon_q+\mu_F}\,.
\ee
This is a leading large $N$ effect: the $1/N$ from the vertices cancels against an $N$ enhancement from the flavors running in the loop.
For all processes of interest, the momentum dependence in the self-energy can be neglected. The reason is that the fermionic momentum scales like $p_\perp \sim \omega$ while the bosonic momentum $q \sim \omega^{1/3}$. Hence $p \ll q$ in the integral above, and $\Sigma$ is approximately independent of $p_\perp$.

Given this, let us parametrize $\vec p = \hat n\,k_F$, and $\vec q = \hat n q_\perp +\vec q_\parallel$, with $\vec q_\parallel$ orthogonal to $\hat n$. The self-energy reads
\be
i \Sigma(\omega_n) = -g^2 T \sum_m \int \frac{d^{d-1} q_\parallel}{(2\pi)^{d-1}}\,\frac{dq_\perp}{2\pi}\,\frac{1}{q_\parallel^2+q_\perp^2+ M_D^2 \frac{|\omega_m-\omega_n|}{\sqrt{q_\parallel^2+q_\perp^2}}}\,\frac{1}{i\omega_m-vq_\perp}\,.
\ee
The usual way to evaluate this integral is to neglect the $q_\perp$ dependence in the boson propagator, since $q_\perp \sim \omega$ while $q_\parallel \sim  \omega^{1/3}$. The two momentum integrals then factorize, obtaining
\be\label{eq:Sigmadiv1}
 \Sigma(\omega_n) = A_d \,\frac{\bar g^2}{v}\,T \sum_m \frac{\text{sgn}(\omega_m)}{(M_D^{-1}|\omega_m-\omega_n|)^{\frac{3-d}{3}}}
\ee
with $\bar g^2 = g^2/M_D^{3-d}$ the dimensionless coupling measured at the Debye scale, and
\be
A_d= \left(\frac{2\pi^\frac{d-1}{2}}{\Gamma(\frac{d-1}{2})} \frac{1}{(2\pi)^{d-1}} \right)\,\frac{\pi}{3 \sin \frac{\pi d}{3}}\,.
\ee

The self-energy (\ref{eq:Sigmadiv1}) is well known in various non-Fermi liquid models, with $2\gamma = \frac{3-d}{3}$ the quasiparticle anomalous dimension. However, it has a basic sickness: it is singular at $m=n$. This is a thermal infrared divergence. This problem is usually ignored in treatments of non-Fermi liquid superconductivity, e.g. appealing to Anderson's theorem; and the sum is restricted to $m \neq n$. However, this is not satisfactory; a consistent QFT should lead to finite well-defined predictions.

The origin of the $m=n$ singularity is in fact the factorization assumption: for $\omega_m \sim \omega_n$, $q_\parallel \sim |\omega_m-\omega_n|^{1/3}$ can be much smaller than $q_\perp$. In this case, the momentum $q_\perp$ carried by the virtual fermion produces a large recoil on the virtual boson, and this effect cannot be neglected. The singularity just reflects the fact that the factorization assumption gives a sum over all $q_\perp$, ignoring the backreaction on the boson. Therefore, factorization breaks down at finite temperature. In contrast, this is a good approximation at zero temperature, where the Matsubara sum is replaced by an integral, and the singularity is integrable.

In order to fix this, let us adopt spherical coordinates $q_\perp^2+q_\parallel^2=q^2$, $q_\perp = q \cos \theta$. Integrating over the angle, and working for simplicity at $d=3-\epsilon$ with $\epsilon \ll 1$, obtains
\be
\Sigma(\omega_n)= 3 \alpha M_D^\epsilon \,T \sum_m\,\int dq \,q^{1-\epsilon}\,D(\omega_m-\omega_n,q)\,\tan^{-1}\left(\frac{v q }{\omega_m} \right)\,.
\ee
We have introduced the dimensionless combination
\be
\alpha = \frac{g^2}{12\pi^2 v} M_D^{-\epsilon}
\ee
that sets the strength of non-Fermi liquid effects. In this form, the thermal singularity at $m=n$, which came from virtual bosons with $q \approx 0$ has been resolved  because the new $\arctan$ factor vanishes there. The calculation now proceeds by splitting the sum into $m \neq n$, and $m=n$, and is done in the Appendix. The final result is
\be\label{eq:Sigmaone}
\Sigma(\omega_n)=\frac{3\alpha}{\epsilon}\,\text{sgn}(\omega_n)\left(  \left(\frac{v M_D}{|\omega_n|} \right)^\epsilon\,\pi T+  M_D^{\epsilon/3}(2\pi T)^{1-\epsilon/3}\left[  \zeta(\epsilon/3)-\zeta(\epsilon/3,|n|+1)\right]\right)\,.
\ee
The first term comes from the resolution of the $m=n$ singularity, while the Riemann zeta functions arise from performing the sum over $m \neq n$. (This last contribution has appeared before in studies of non-Fermi liquids \cite{abanov2003quantum, Moon2010}.)

Let us analyze the low and high temperature limits. To access low $T$ we take large $n$, for which $\zeta(\epsilon/3)-\zeta(\epsilon/3,n+1) \sim n^{1-\epsilon/3}$. In this limit,
\be
\Sigma(\omega_n) \approx \frac{3\alpha}{\epsilon} M_D^{\epsilon/3} |\omega_n|^{1-\epsilon/3}\,\text{sgn}(\omega_n)\,.
\ee
This recovers the $T=0$ result of \cite{Torroba:2014gqa, Raghu:2015sna}, reviewed above in (\ref{eq:GFQCP}). In this regime we have a metallic QCP with fermion anomalous dimension $2\gamma= \epsilon/3$. The non-Fermi liquid scale (the scale at which the self-energy dominates over the tree level term) is hence
\be\label{eq:LNFL}
\Lambda_{NFL} \approx \left(\frac{3\alpha}{\epsilon} \right)^{3/\epsilon}\,M_D\,.
\ee
At weak coupling, $\alpha<\epsilon/3$, and $\Lambda_{NFL}\ll M_D$. On the other hand, for $\alpha=\epsilon/3$, $\Lambda_{NFL}\sim M_D$, and one starts from the QCP at the UV scale.

At high temperatures, but still well below the microscopic scale $M_D$, the first term in (\ref{eq:Sigmaone}) dominates over the $\zeta$ functions. In particular, we obtain a new thermal contribution at the first Matsubara frequency
\be
\Sigma(\pi T) \approx \frac{3 \alpha}{\epsilon} (\pi T)^{1-\epsilon} (v M_D)^\epsilon\,.
\ee
This IR result is parametrically larger than the NFL answer by a factor $(M_D/T)^{2\epsilon/3} \gg 1$.

%%%%%%%%%%%%%%%%%%%%%%%%%%%%%%%%%%%%%%%
%%%%%%%%%%%%%%%%%%%%%%%%%%%%%%%%%%%%%%%
\subsection{An infrared catastrophe}\label{subsec:catastrophe}

Although it seems we have solved the IR problems, unfortunately this is not the case --another type of singularity is hiding in (\ref{eq:Sigmaone}). The reason is that it is not sufficient to obtain a finite self-energy over the Matsubara frequencies; one also needs a sensible result on the upper half complex plane, which determines the dynamics in real time.

To see this, recall that the retarded Green's function (which e.g. enters transport calculations) is related to the Euclidean result by
\be
i \omega_n \to p_0+i \delta\;,\;\delta \to 0^+
\ee
and
\be
i \omega_n+i \Sigma_E(\omega_n) \to p_0+i \Sigma_E(-i p_0+\delta) \equiv  p_0 + \Sigma_{ret} (p_0)\,.
\ee
Let us then perform the analytic continuation of the new thermal term in (\ref{eq:Sigmaone}). We have $\omega_n^2 \to (-ip_0+\delta)^2= e^{-i \pi \text{sgn}(p_0)} p_0^2$, from which we deduce that $|\omega_n| \to -i p_0$, and $\text{sgn}(\omega_n) = \frac{\omega_n}{|\omega_n|} \to +1$. Thus, the high temperature euclidean result gives a retarded self-energy
\be\label{eq:retone}
\Sigma_{ret}(p_0) \approx i \frac{3\alpha}{\epsilon}\, \left(\frac{v M_D}{|p_0|} \right)^\epsilon\,e^{i\frac{\pi}{2}\epsilon\,\text{sgn}(p_0)}\pi T\,.
\ee
This gives a decay rate that diverges as $p_0 \to 0$! In particular, the perturbative expansion always breaks down at sufficiently low energies.

The physical origin of this divergence was described in \S \ref{sec:intro}: at low frequencies, the Bose-Einstein occupation factor leads to powers of $T/\omega$, as in (\ref{eq:IRdiv}). To see this more explicitly, let us go back to (\ref{eq:Sigmaone0}), introduce the spectral density
\be
D(\omega_n, q) = \int_{-\infty}^\infty \frac{dq_0}{2\pi}\,\frac{\rho_B(q_0,q)}{q_0-i \omega_n}
\ee
and perform the Matsubara sum to obtain 
\be
i \Sigma(\omega_n)= - g^2  \int_{-\infty}^\infty \,\frac{d q_0}{2\pi} \,\int \frac{d^d q}{(2\pi)^d}\, \rho_B(q_0, q)\,\frac{n_F(\varepsilon_q)+n_B(-q_0)}{\varepsilon_q+q_0-i\omega_n}	\,.
\ee
Here $n_F$ and $n_B$ are the Fermi-Dirac and Bose-Einstein distributions. (More details on this and the following calculations are discussed in the Appendix.) 

Now, at low frequencies $n_B(q_0) \sim T/q_0$, and in this case (for momenta well below $M_D$)
\be\label{eq:density}
n_B(q_0) \rho_B(q_0, q) \approx 2\pi T \frac{\delta(q_0)}{q^2}\,.
\ee
This means that the density of bosonic states is concentrated at zero frequency. This aspect of overdamped bosons has been recognized in works on QED and QCD at finite density (see e.g. \cite{Blaizot:1996az} and references therein). As we are now finding, it also plays a key role in quantum criticality. Using these properties to calculate the imaginary part of the retarded self-energy leads to the same result (\ref{eq:retone}).

To summarize, the coupling of a Fermi surface to a Landau damped critical boson leads to finite temperature divergences that come from exchanging virtual soft bosons whose density of states is peaked at zero frequency. Moreover, each addition of a virtual boson makes the result more infrared singular. The perturbative expansion  breaks down, and in order to determine the properties of the QCR it is essential to understand how to cure this problem.

%%%%%%%%%%%%%%%%%%%%%%%%%
%%%%%%%%%%%%%%%%%%%%%%%%%
%%%%%%%%%%%%%%%%%%%%%%%%%
%%%%%%%%%%%%%%%%%%%%%%%%%
\section{The jump to all orders: rainbow resummation}\label{sec:nonpert}

The resolution of infrared singularities is an important open problem in finite temperature QFT,\cite{le2000thermal} and it generally requires nonperturbative methods. One logical possibility is that the IR phase is gapped, and then this mass scale enters as an IR regulator that avoids the singularities. For instance, one could imagine a thermal mass for the boson, or that superconductivity gaps the Fermi surface. However, this approach is not always correct. For instance, in the analog problem of an emergent gauge field (instead of a scalar field), gauge invariance forbids a mass term and hence a different mechanism is required. Moreover, assuming that the divergences are cut off by the superconducting gap leads to a singular dependence in the gap equation that is unphysical. Instead, our view here is that the theory should cure its own singularities, without appealing to new physics

\begin{figure}[h!]
\centering
\includegraphics[width=0.9\textwidth]{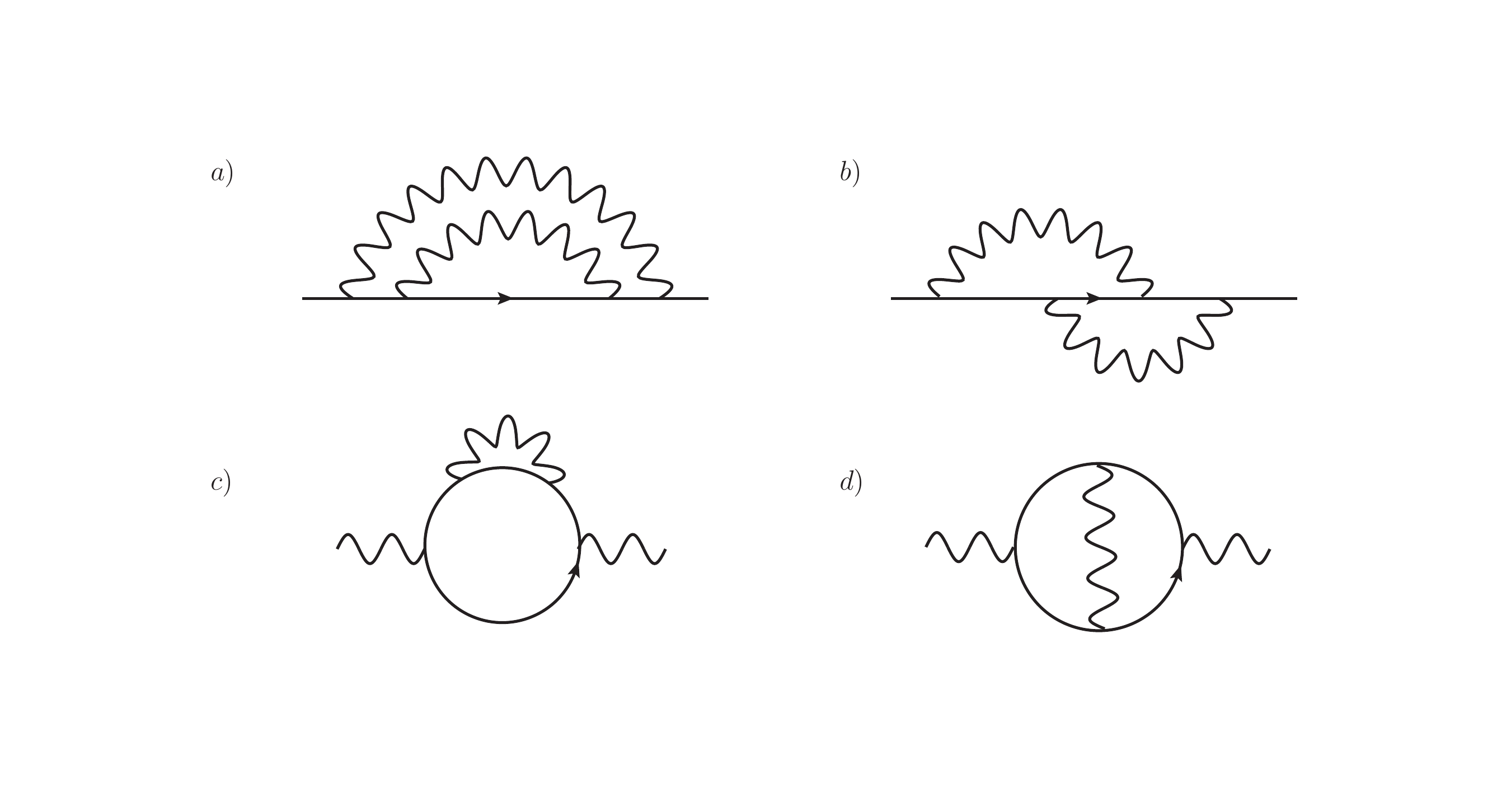}
\caption{Two-loop corrections to the fermion and boson self-energy. b) and d) are subleading at large $N$.}\label{fig:2loop}
\end{figure}

To continue along this direction, we need to understand what happens at higher loops. Fig. \ref{fig:2loop} shows a few two-loop diagrams, where the addition of virtual boson lines leads to higher order IR singularities. The idea is that resumming quantum corrections should lead to finite well-defined results, however, this is not tractable in general and some truncation is needed. Along these lines, works on QED and QCD at finite density addressed this problem in terms of the Bloch-Nordsiek resummation\cite{Blaizot:1996az}; Ref.\cite{abanov2003quantum} also employed methods similar to ours in the context of antiferromagnetic models. The main point is that the large $N$ limit produces a dramatic simplification of quantum corrections: only rainbow diagrams (such as those in a) and c) of Fig. \ref{fig:2loop}) survive, while diagrams that contain vertex corrections as subdiagrams (b) and d) of Fig. \ref{fig:2loop}) are suppressed by powers of $1/N$. Therefore, the resolution of thermal divergences has to come from resumming the rainbow diagrams. 

\begin{figure}[h!]
\centering
\includegraphics[width=1.\textwidth]{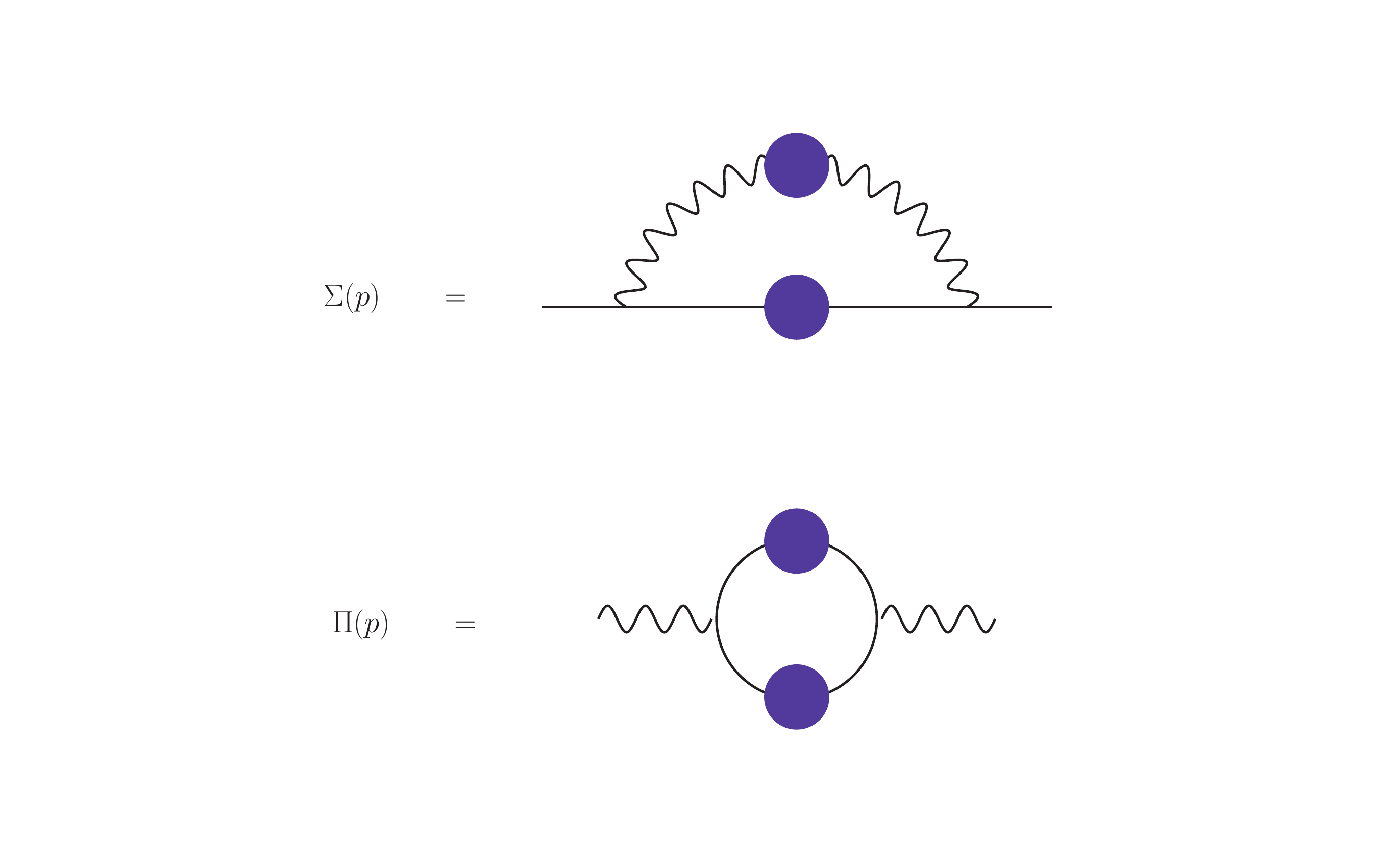}
\caption{Schwinger-Dyson equations for the fermion and boson self-energy in the large $N$ theory.}\label{fig:SDeq}
\end{figure}

This is encoded in the following coupled set of Schwinger-Dyson equations for the bosonic and fermions self-energies:
\bea\label{eq:SD}
\Pi(\omega_n, q) &=& \frac{g^2}{N}\, T \sum_m \int\frac{d^d p}{(2\pi)^d}\,G(\omega_m, p)\,G(\omega_m+\omega_n, p+q) \nonumber\\
i \Sigma( \omega_n ,q)&=& g^2 T \sum_m \int\frac{d^d p}{(2\pi)^d}\, D(\omega_m-\omega_n, p-q) G(\omega_m, q)\,.
\eea
Here $D, G$ denote the full propagators defined in (\ref{eq:fullGs}). The diagrammatic interpretation is described in Fig. \ref{fig:SDeq}. In this section we will solve these equations and show how the IR problems are cured.

%%%%%%%%%%%%%%%%%%%%%%%%%%%%%%%%%%%%%%%
%%%%%%%%%%%%%%%%%%%%%%%%%%%%%%%%%%%%%%%
\subsection{Landau damping including quantum and thermal corrections}\label{subsec:rainbow-Pi}

Let us analyze the quantum corrections to the boson propagator. At zero temperature, $\Pi(\omega, q)$ is one loop exact in the $z=3$ scaling regime for the boson\cite{Polchinski1994} (we also review below how this comes about). Here we are interested in the result at finite temperature.

Focusing for simplicity on $d=3$, and linearizing $\varepsilon_p \approx v_F p$ near the Fermi surface, we have
\be
\Pi(i \omega_n, q) = \frac{g^2 k_F^2}{(2\pi)^2} \,\int_{-1}^1 d\cos \theta\,\int dp_\perp\,\frac{1}{\beta} \sum_m\, G(i\omega_m, p_\perp) G(i \omega_m+ i \omega_n, p_\perp+q \cos\theta)\,.
\ee
Here $\omega_n$ is a bosonic Matsubara frequency, while $\omega_m$ is fermionic.
Performing the momentum integral by residues obtains
\be
\Pi(\omega_n, q) =-\frac{g^2 k_F^2}{(2\pi)^2 v} \int_{-1}^1 dx\,\frac{2\pi i}{\beta} \sum_m\,\frac{\Theta(\omega_m+\omega_n)-\Theta(\omega_m)}{v q x - i \omega_n -i \left(\Sigma(\omega_m+\omega_n)-\Sigma(\omega_m) \right)}
\ee
Note that $\Pi(0,q)=0$. Integrating over the angle and taking into account the restrictions from the step functions,
\be\label{eq:Pithermal}
\Pi(\omega_n, q) =\left(\frac{g^2 k_F^2}{2\pi^2 v}\right) \frac{1}{q v}\frac{2\pi}{\beta} \sum_{m=0}^{|n|-1}\,\tan^{-1}\left(\frac{q v}{\omega_{|n|} + \Sigma(\omega_{|n|}-\omega_m)+\Sigma(\omega_m)}\right)\,.
\ee
This is our main result for the Landau damping function that includes thermal and quantum effects. The one loop answer (\ref{eq:Pi1}) is recovered by setting $\Sigma(\omega)=0$.

The boson dynamics at one loop is governed by the $z=3$ scaling $q^3 \sim M_D^2 \omega$ discussed in \S \ref{subsec:Pi1}, and we want to see how this is modified by (\ref{eq:Pithermal}). A more detailed analysis is included in the appendix. Consider first the low temperature behavior $T \ll \omega_n$. Since $\Sigma(\omega) = \Lambda_{NFL}^{2\gamma} \omega^{1-2\gamma}$, the inverse tan function is, for $q\sim M_D^{2/3}\omega^{1/3}$, of order
\be
\tan^{-1}\left(\frac{v q}{ \Sigma( \omega)}\right) \sim \tan^{-1}\left(\left(\frac{M_D}{\Lambda_{NFL}}\right)^{2\gamma} \left(\frac{M_D}{\omega}\right)^{\frac{2}{3}-2\gamma} \right) \approx \frac{\pi}{2}
\ee
at low energies. We assumed here that $\gamma<\frac{1}{6}$, which is certainly the case for $\epsilon<1$. Therefore, due to the $z=3$ scaling, the Landau damping function is independent of the fermion self-energy -- it coincides with the one loop result to all orders in perturbation theory. 

For low frequencies $\omega_n \sim \pi T$ we will argue shortly that there appears a new contribution to the fermion self-energy of the form $\Sigma (\pi T) \sim (M_D^\epsilon \pi T)^{\frac{1}{1+\epsilon}}$. This could in principle modify the behavior of (\ref{eq:Pithermal}) for small bosonic Matsubara frequency, i.e. $\omega_n \sim 2\pi T$. Making use of the scaling $q^3 \sim M_D^2 \omega \sim M_D^2 \pi T$, in this regime one finds
\be\label{eq:Pistrong}
\tan^{-1}\left(\frac{v q}{ \Sigma (\pi T)}\right) \sim \tan^{-1}\left[\left( \frac{M_D}{\pi T}\right)^{\frac{1-2\epsilon}{3(1+\epsilon)}} \right]\,.
\ee
Since $\pi T \ll M_D$ and we are working at small $\epsilon$, this is again of order $\pi/2$ and hence Landau damping is not changed. Curiously, this can modify appreciably the boson Landau damping for $\epsilon>1/2$, which is of interest for $d=2$ physics. Nevertheless, the large $N$ approximation is also expected to break down at strong coupling \cite{Lee2009}, so at this stage the relevance of such thermal corrections near $d=2$ is not clear.

%%%%%%%%%%%%%%%%%%%%%%%%%%%%%%%%%%%%%%%
%%%%%%%%%%%%%%%%%%%%%%%%%%%%%%%%%%%%%%%
\subsection{Fermion self-energy to all orders}\label{subsec:rainbow-Sigma}

Having understood the absence of qualitative modifications to $\Pi\left(\omega_m,q\right)$ from $\Sigma(\omega_n)$, we can now use the $z=3$ representation for $D\left(\omega_n,q\right)$ and focus on the self-consistent Schwinger-Dyson equation for $\Sigma\left(\omega_n\right)$:
\bea
i \Sigma( \omega_n)&=&-g^2 T \int\frac{d^d p}{(2\pi)^d} D\left(0,p\right)\frac{1}{i\omega_n+vp_\perp +i\Sigma\left(\omega_n\right)}\nonumber\\
&-& g^2 T \sum_{m\neq n}\int\frac{d^d p}{(2\pi)^d} D\left(\omega_n-\omega_m,p\right)\frac{1}{i\omega_m+vp_\perp +i\Sigma\left(\omega_n\right)}\,,
\eea
 where again we separate the Matsubara sum into a thermal term $\omega_m=\omega_n$ and the remaining dynamical contributions $\left(\omega_m\neq \omega_n\right)$. The latter can be approximated using momentum factorization due to $z=3$ scaling and is independent of $\Sigma\left(\omega_m\right)$; evaluating the thermal term as before using spherical coordinates for $p$ lands us on the following self-consistent equation:
\begin{eqnarray}\label{eq:self-energy-gen}
&&\Sigma(\omega_n) = \frac{3\alpha}{\epsilon} M_D^\epsilon \left(c_1 \pi T \left[\omega_n+\Sigma(\omega_n)\right]^{-\epsilon}+c_2 (2\pi T)^{1-\frac{\epsilon}{3}} \frac{\zeta(\epsilon/3)-\zeta(\epsilon/3,n+1)}{M_D^{2\epsilon/3}}\right),\;\;\omega_n>0\nonumber\\
&& c_1 =\frac{v^\epsilon}{\pi}\Gamma\left(\frac{1-\epsilon}{2}\right)\Gamma\left(\frac{1+\epsilon}{2}\right),\;c_2= \Gamma\left(1-\frac{\epsilon}{3}\right)\Gamma\left(1+\frac{\epsilon}{3}\right)
\end{eqnarray}

The first term in (\ref{eq:self-energy-gen}) encodes the resummation of rainbow diagrams, giving a self-consistent equation for the self-energy. We will see that this resolves the infrared singularities encountered before in perturbation theory. Note that this term is in fact the only contribution at the first Matsubara frequency $\omega=\pm \pi T$, because the two Riemann zeta function terms cancel out for $n=0, -1$. This property of the first Matsubara frequency plays a special role in the NFL dynamics discussed below.

It is easy to solve (\ref{eq:self-energy-gen}) numerically, and one obtains a fully nonsingular fermion self-energy. We show this below in Fig. \ref{fig:self-energy}. Nevertheless, it is useful to develop an approximate solution where quantum and thermal effects can be seen more explicitly. For this, we note that at high frequencies  the dynamical terms dominate over the thermal term, and the self energy is simply controlled by the NFL regime: 
\begin{equation}
\Sigma(\omega_n)\sim  c_2\,\frac{3\alpha}{\epsilon} M_D^{\epsilon/3}|\omega_n|^{1-\epsilon/3}\sgn(\omega_n)\,.
\end{equation}
On the other hand, the thermal term dominates at low frequencies, and then the self-energy is controlled by the constant solution of the algebraic equation
\begin{equation}\label{eq:algebraic}
\Sigma(\pi T) \approx \frac{3\alpha}{\epsilon}c_1\pi T\,M_D^\epsilon |\pi T +\Sigma(\pi T)|^{-\epsilon}\,.
\end{equation}
We denote the solution by $\Sigma_T$.
The low and high temperature behaviors can be obtained iteratively, finding
\be\label{eq:SigmaT}
\Sigma(\pi T)=\Sigma_T \approx \left \lbrace \begin{matrix}
 \left(\frac{3\alpha}{\epsilon}c_1\,M_D^\epsilon \pi T\right)^{\frac{1}{1+\epsilon}}\;,& \frac{\pi T}{M_D} \ll \left( \frac{\Lambda_{NFL}}{M_D}\right)^{1/3} \\ 
\frac{3\alpha}{\epsilon}c_1\,M_D^\epsilon (\pi T)^{1-\epsilon}\;,& \frac{\pi T}{M_D} \gg \left( \frac{\Lambda_{NFL}}{M_D}\right)^{1/3}
  \end{matrix} \right. \,.
\ee
We will denote the low and high temperature branches by $\Sigma_T^{(1)}$ and $\Sigma_T^{(2)}$, respectively.
The full self-energy is then approximated by
\be\label{eq:Sigma-final}
\sgn(\omega_n)\Sigma(\omega_n)\approx \Sigma_T+c_2 \frac{3\alpha}{\epsilon} M_D^\epsilon  (2\pi T)^{1-\frac{\epsilon}{3}} \frac{\zeta(\epsilon/3)-\zeta(\epsilon/3,n+1)}{M_D^{2\epsilon/3}}\,.
\ee
This is our main result for the exact fermion self-energy in the large $N$ theory.

In Fig \ref{fig:self-energy} we plot the approximate analytic solution against the numerical solution to (\ref{eq:self-energy-gen}) -- the agreement is excellent. We also show the temperature dependence of $\Sigma(\pi T)$ for different values of $\epsilon$. In particular, note that the low and high temperature limits are quite different as $\epsilon$ is increased. Even though it is not clear whether the framework remains under control at strong coupling, it is interesting to extrapolate these results to $\epsilon =1$, corresponding to a model in $d=2$. In this case, the low temperature result gives $\Sigma(\pi T) \sim T^{1/2}$, while at high temperatures $\Sigma(\pi T)$ becomes independent of $T$.\footnote{More precisely, there is a subleading $\log T$ dependence from expanding (\ref{eq:SigmaT}) near $d=2$.} It would be interesting to understand the effects of this on e.g. transport properties, a point which we hope to study in the future.
\begin{figure}[h!]
\centering
\includegraphics[width=0.50\textwidth]{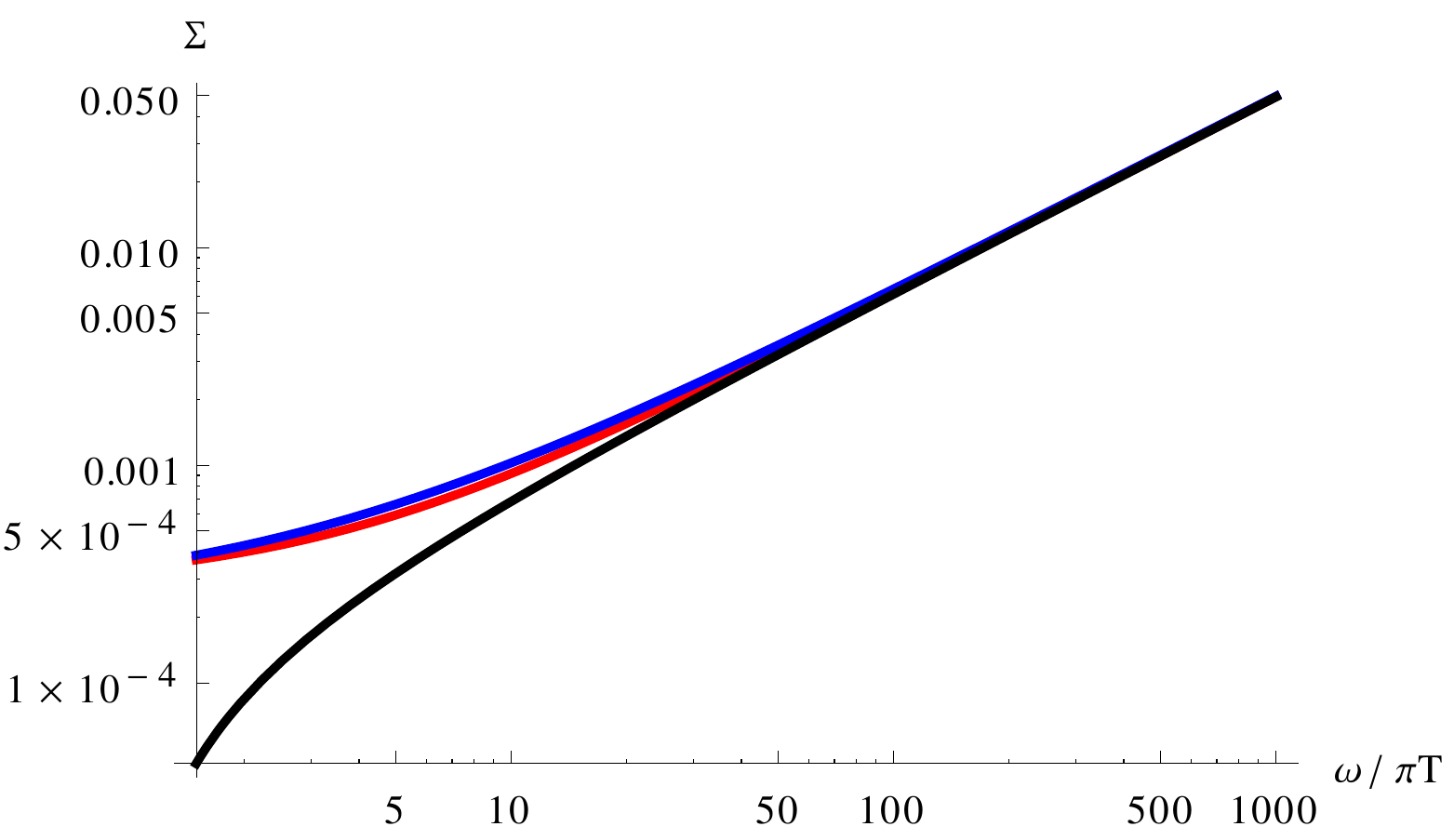}
\includegraphics[width=0.48\textwidth]{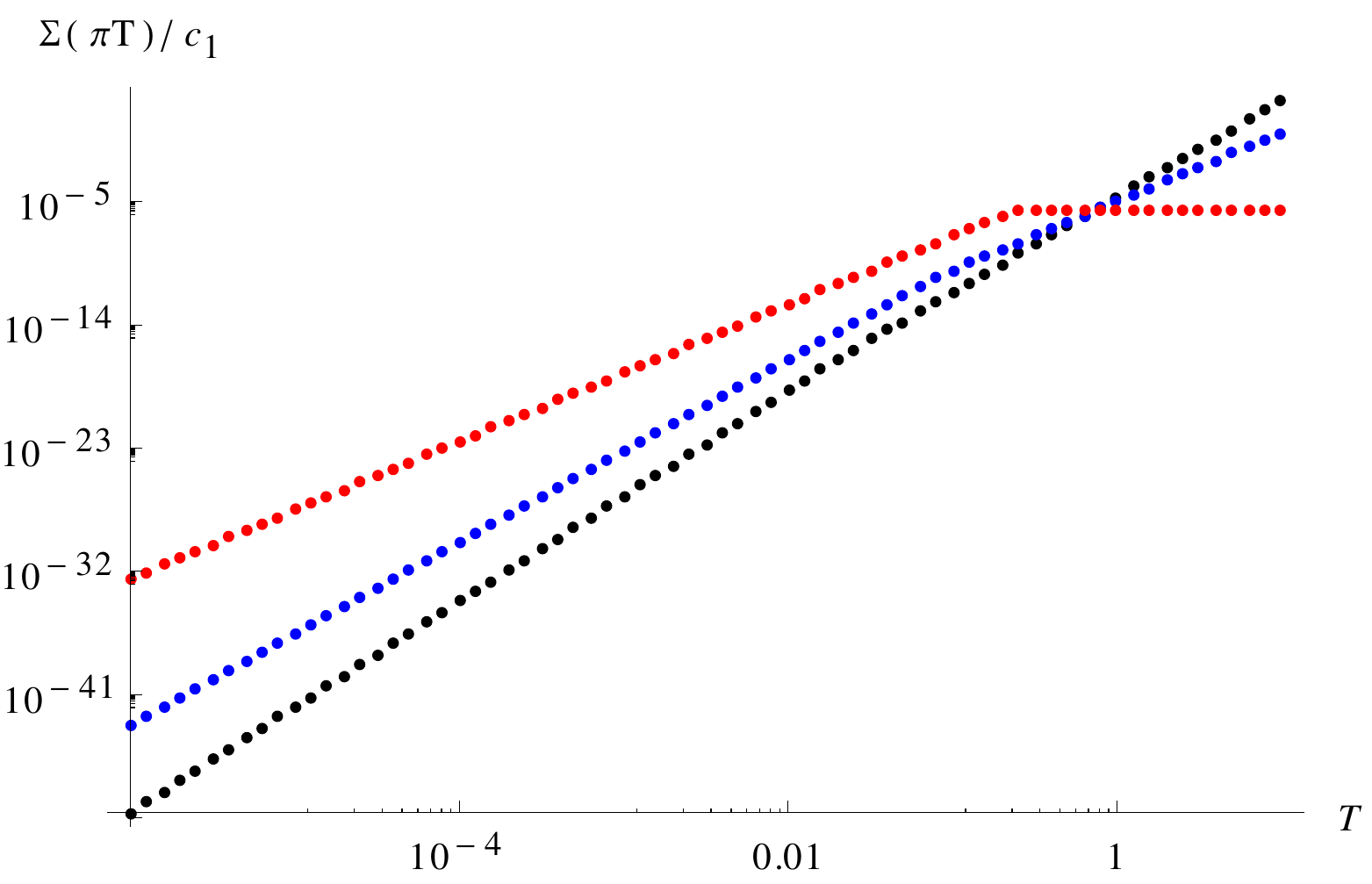}
\caption{Left panel:  exact solution to (\ref{eq:self-energy-gen}) (red), approximate solution (\ref{eq:Sigma-final}) (blue), and NFL answer ignoring the thermal piece (black). Parameters: $\epsilon=0.3, \alpha=\epsilon/3, \pi T/M_D= 10^{-5}$. Right panel: evolution of $\Sigma(\pi T)$ as a function of $T$ for $\epsilon=0.3$ (black), $\epsilon=0.5$ (blue), $\epsilon=1$ (red); we fix $M_D=1$. }\label{fig:self-energy}
\end{figure}

We conclude that in the class of NFLs considered in this work, the exact thermal Green's functions can be calculated in the large $N$ limit, and that this resummation resolves the IR singularities encountered within perturbation theory. This gives rise to a new thermal piece in the self-energy, which is approximately independent of the external frequency and dominates in the static limit. Note that even though this is a high temperature effect ($\omega \sim \pi T$), it is highly nonperturbative; it encodes the NFL nature of the Fermi surface in the nontrivial power-law dependence on temperature.

%%%%%%%%%%%%%%%%%%%%%%%%%
%%%%%%%%%%%%%%%%%%%%%%%%%
%%%%%%%%%%%%%%%%%%%%%%%%%
\section{Implications and discussion}\label{sec:implications}

We are now ready to explain the NFL behavior in the quantum critical region. At $T=0$ we have two scales: $M_D$ (the UV cutoff) and the NFL scale $\Lambda_{NFL}$ introduced in (\ref{eq:LNFL}). This is controlled by the ratio $3\alpha/\epsilon$, with $\alpha \ll \epsilon/3$ the weak coupling limit. On the other hand, for $T \neq 0$ a richer dynamics emerges in the QCR.

To begin, it is necessary to distinguish between two types of NFL behavior. In the \textit{quantum} NFL regime, the $T=0$ self-energy dominates the dynamics, $|\omega_n| \ll \Lambda_{NFL}^{\epsilon/3} |\omega_n|^{1-\epsilon/3}$. On the other hand, we define the \textit{thermal} NFL regime as the case when the thermal term in the self-energy dominates the dynamics, $\omega \ll \Sigma_T$. In particular, for $\pi T > \Lambda_{NFL}$, the quantum NFL self energy $\Sigma \sim \Lambda_{NFL}^{\epsilon/3}|\omega_n|^{1-\epsilon/3}$ is negligible, however, the thermal term $\Sigma_T$ can still dominate over a range of frequencies. Note also that the temperature dependence in the thermal NFL case is given by a nontrivial power-law, and does not obey mean-field scaling. Furthermore, scaling arguments that are used near QCPs at finite $T$ are violated in the thermal NFL regime.

Let us explore the QCR starting from the highest temperatures. By comparing the different terms in the full fermionic Green's function we obtain the following regimes (see the Appendix):

$\bullet$ $\pi T > \Lambda_{NFL}^{1/3} M_D^{2/3}$. All quantum and thermal corrections are subdominant, leading to Fermi liquid (FL) behavior.

$\bullet$ $\left(\frac{\Lambda_{NFL}}{M_D} \right)^{2\epsilon/3}\Lambda_{NFL}< \pi T < \Lambda_{NFL}^{1/3} M_D^{2/3}$. Quantum corrections are subdominant, but the thermal term in $\Sigma$ can be important. In this range, the high frequency limit is a FL, while at low frequencies a thermal NFL emerges, dominated by the low temperature branch of (\ref{eq:SigmaT}),
\be
\omega+\Sigma(\omega) \approx \left \lbrace\begin{matrix}\omega\;, &  \omega \gg \Sigma_T^{(1)}\\ \Sigma_T^{(1)}=\left(\Lambda^{\epsilon/3}_{NFL} M_D^{2\epsilon/3}\pi T\right)^{\frac{1}{1+\epsilon}}\;,  & \omega \ll \Sigma_T^{(1)} \end{matrix}  \right. \,.
\ee
For $\alpha \sim \epsilon/3$ or for sufficiently low temperatures, we thus find a new NFL thermal regime over a range of frequencies $\omega< \Sigma_T^{(1)}$.

$\bullet$ $\pi T <\left(\frac{\Lambda_{NFL}}{M_D} \right)^{2\epsilon/3}\Lambda_{NFL}$. In this range one finds all the possibilities for the fermion Green's function: FL, quantum NFL, and thermal NFL:
\be
\omega+\Sigma(\omega) \approx \left \lbrace\begin{matrix}\omega\;, &  \omega \gg \Lambda_{NFL}\\ \Lambda_{NFL}^{ \epsilon/3} \omega^{1-\epsilon/3}\;,  & \Lambda_T \ll \omega \ll \Lambda_{NFL}\\ \Sigma_T^{(1)}\;,& \omega \ll \Lambda_T \end{matrix}  \right.
\ee
with cross-over scale
\be
\Lambda_T \sim \left( \frac{\Sigma_T^{(1)}}{\Lambda_{NFL}^{\epsilon/3}}\right)^{\frac{1}{1-\epsilon/3}}\,.
\ee
We stress that the thermal NFL behavior cannot be obtained from the QCP at $T=0$ --it is a purely thermal effect. The thermal piece always dominates at the first Matsubara frequency $\omega= \pi T$.

\begin{figure}[h!]
\centering
\includegraphics[width=0.7\textwidth]{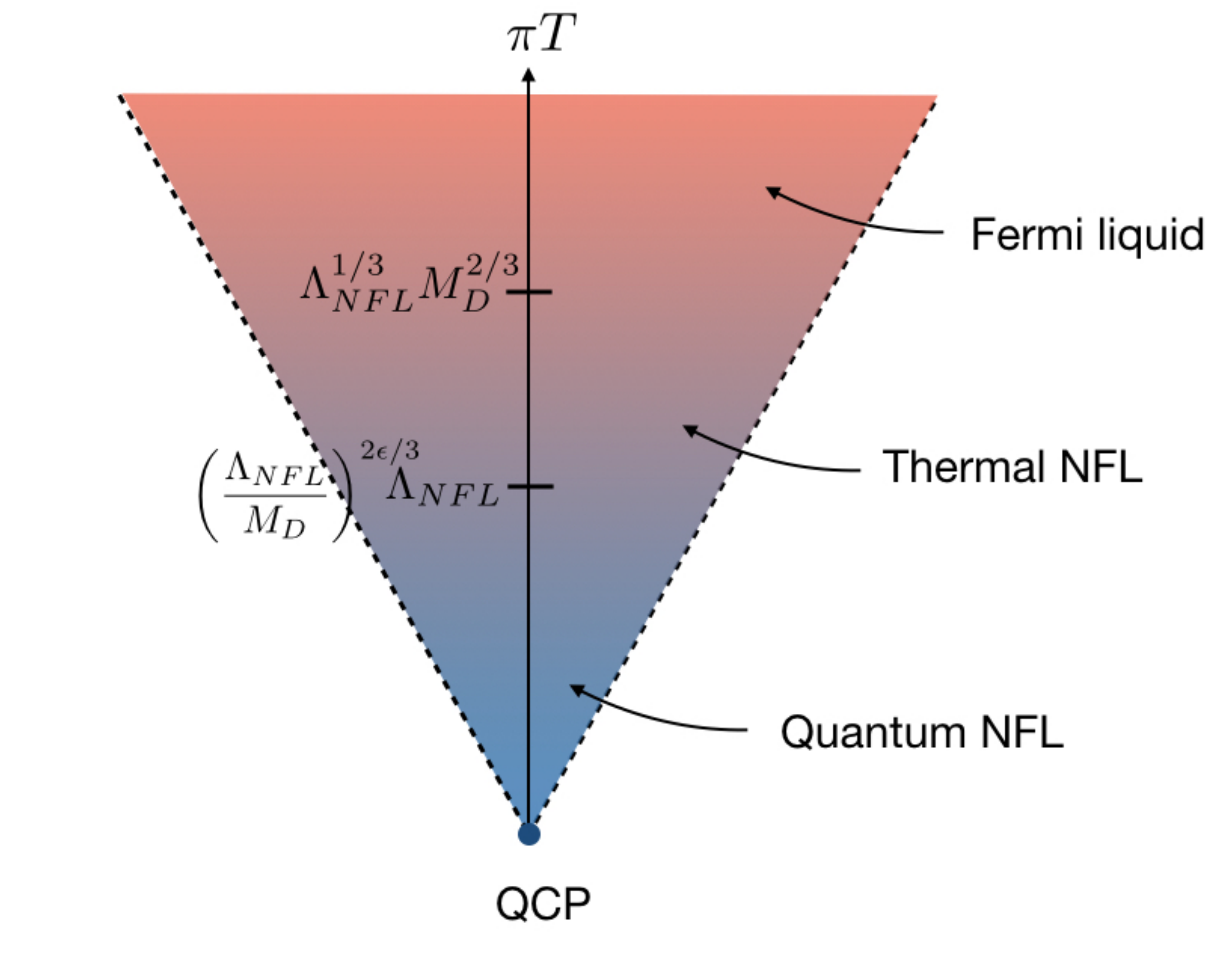}
\caption{Phase structure of the quantum critical region near a NFL fixed point.}\label{fig:phases}
\end{figure}

We summarize the resulting phase structure, together with the cross-over scales, in Fig. \ref{fig:phases}. Each regime is labelled by the new behavior that appears there. As discussed before, in each range one has to distinguish between static and dynamic properties of the fermionic Green's function.

Although most of our analysis has been restricted to euclidean time, we expect that the thermal NFL regime will have a nontrivial impact on real time observables. As a first step, let us now calculate the real time self-energy. For this, we analytically continue the full euclidean answer as discussed in \S \ref{subsec:catastrophe}. The constant thermal term gives rise to a purely imaginary contribution, while the quantum NFL piece translates into both real and imaginary parts:
\be
\Sigma_{ret}(p_0)= i\left( \frac{3\alpha}{\epsilon}c_1\, \pi T M_D^\epsilon \right)^{\frac{1}{1+\epsilon}}+i\,c_2\,\frac{3\alpha}{\epsilon} M_D^{\epsilon/3}(2\pi T)^{1-\epsilon/3}\left[\zeta(\epsilon/3)-\zeta\left(\epsilon/3, -i\frac{p_0}{2\pi T} +1\right)\right]\,.
\ee
The dependence at low and high temperatures is
\be\label{eq:Sigmaret2}
\Sigma_{ret}(p_0)\approx \left \lbrace \begin{matrix} i\left( \frac{3\alpha}{\epsilon}c_1\, \pi T M_D^\epsilon \right)^{\frac{1}{1+\epsilon}}+c_2\,\frac{3\alpha}{\epsilon} M_D^{\epsilon/3}p_0 \, e^{-i \frac{\pi}{2}\text{sgn}(p_0) \frac{\epsilon}{3}} \,|p_0|^{-\epsilon/3}\;,&\pi T \ll |p_0| \\ 
i\left( \frac{3\alpha}{\epsilon}c_1\, \pi T M_D^\epsilon \right)^{\frac{1}{1+\epsilon}}+c_2\frac{3\alpha}{\epsilon} M_D^{\epsilon/3} (2\pi T)^{-\epsilon/3}p_0\;,& \pi T \gg |p_0|
  \end{matrix} \right. \,.
\ee
A direct calculation in terms of the spectral density and the Kramers-Kronig relation agrees with these results.

While in the euclidean approach both quantum and thermal terms are on the same footing, in real time they have quite different effects. The quantum term contributes dominantly to the real part of the self-energy, while the thermal term amounts to a NFL decay rate of order
\be\label{eq:tau}
\frac{1}{\tau} \sim \left(\frac{M_D}{T} \right)^{-2\epsilon/3} T \,.
\ee
Note that $1/\tau \gg T$ due to non-Fermi liquid effects. Since the finite $T$ scaling of the QCP is invalidated by $\Sigma_T$, we expect important effects on transport properties. This is a topic that we hope to study in the future.

More generally, it would be interesting to explore thermal NFL signatures in other models. As we argued in this work, such effects should occur whenever IR singularities arise perturbatively. Potential candidates for this could be\cite{meszena2016nonperturbative, saterskog2016exact, lee2017recent}.

Finally, the main future step will be to add 4-fermion interactions along the BCS channel, and study the competition between NFL behavior and superconductivity. The analysis done in Ref. \cite{Wang:2016hir} at $T=0$ showed that at small $N$ the QCP is unstable to superconductivity, but that as $N$ is increased, superconducting fluctuations become more incoherent and dominated by NFL effects. For $N>12/\epsilon$, the BCS coupling becomes critical and superconductivity disappears. This is a naked QCP. Given the results in the present work, it is natural to ask how this picture is modified at finite $T$. Here we just point out that the constant thermal term in the self-energy amounts to an anomalous dimension that increases like $\gamma \sim 1/\omega$ at low frequencies. The BCS beta function constructed in \cite{Wang:2016hir} then shows that the BCS coupling becomes more irrelevant than at $T=0$. Furthermore, the thermal term dominates at the first Matsubara frequency, as noted above, so we do not expect the ``first Matsubara law'' of\cite{2016PhRvL.117o7001W} to operate in this context. Overall, the thermal NFL contribution goes in the direction of destroying superconductivity and increasing the strength of NFL effects in the QCR. This analysis will appear in\cite{future}. 

%%%%%%%%%%%%%%%%%%%%%%%%%
%%%%%%%%%%%%%%%%%%%%%%%%%
%%%%%%%%%%%%%%%%%%%%%%%%%
\acknowledgments{We gratefully thank S. Raghu and Y. Wang for collaboration on related topics, past and on-going discussions on NFLs, and comments on the manuscript. G.T. is supported by CONICET, ANPCYT PICT grant 2015-1224, and CNEA. H.W. is supported by DARPA YFA contract D15AP00108. }

\appendix

\section{Some explicit calculations}

This appendix contains explicit calculations for some of the results quoted in the main text.

\subsection{One loop self-energy}

The one loop fermion self-energy is given by
\be\label{eq:SD-euclidean1}
\Sigma(\omega_n) = 6 \alpha T \sum_{m=0}^\infty\,\int_0^\infty\,q^{1-\epsilon}dq\,\left(D(i \omega_m-i\omega_n,q)-D(i\omega_m+i\omega_n,q) \right)\,\tan^{-1}\left(\frac{v q}{\omega_m}\right)\,.
\ee
In this form, it is clear that $\Sigma(-\omega_n) = -\Sigma(\omega_n)$. So it is sufficient to consider positive frequencies.

To proceed, we will propose some approximations in the momentum-dependence, so that the momentum integral can be performed explicitly.
Recall that
\be
D(i \omega_n, q) = \frac{1}{\omega_n^2+q^2+\frac{2 M_D^2}{\pi}\frac{\omega_n}{q}\tan^{-1}\frac{vq}{\omega_n}}\,.
\ee
For a given Matsubara frequency $\omega_m$ in (\ref{eq:SD-euclidean1}), we split the momentum integral into $vq< \omega_m$ and $vq >\omega_m$. In each range, the inverse tan function and boson propagator simplify as follows:
\bea
\int_0^\infty\,q^{1-\epsilon}dq\,D(i \omega_m-i\omega_n,q)\,\tan^{-1}\left(\frac{v q}{\omega_m}\right)& \approx& \int_0^{\omega_m/v}\,dq\,\frac{1}{|\omega_m-\omega_n|^2+q^2+ \frac{2}{\pi}M_D^2 v}\,\frac{v q^{2-\epsilon}}{\omega_m}\nonumber\\
&+&\frac{\pi}{2} \int_{\omega_m/v}^\infty \,q^{1-\epsilon} dq\,\frac{1}{q^2+M_D^2  \frac{|\omega_m-\omega_n|}{q}}
\eea
for $m \neq n$.  The first part is suppressed by the large Debye scale $M_D$, and we may neglect it in our low energy EFT. The second part is the usual $z=3$ contribution; hence
\be\label{eq:semi-analytic}
\int_0^\infty\,q^{1-\epsilon}dq\,D(i \omega_m-i\omega_n,q)\,\tan^{-1}\left(\frac{v q}{\omega_m}\right) \approx  \frac{\pi}{2\epsilon}\,\frac{1 }{|\omega_m-\omega_n|^{\epsilon/3}}
\,.
\ee

The only special case is $m=n$, which may be evaluated explicitly:
\be
\int_0^\infty\,q^{1-\epsilon}dq\,D(0,q)\,\tan^{-1}\left(\frac{v q}{\omega_n}\right)=\frac{\pi}{2\epsilon}\,\frac{v^\epsilon}{\omega_n^\epsilon}\,.
\ee
Putting everything together,
\be
\Sigma(\omega_n) \approx 6 \alpha T \left[ \frac{\pi}{2\epsilon}\,\frac{v^\epsilon}{\omega_n^\epsilon} - \frac{\pi}{2\epsilon}\,\frac{1}{(2\omega_n)^{\epsilon/3}}\right]+\frac{3 \alpha}{\epsilon}\pi T \sum_{m=0}^\infty \left(\frac{1-\delta_{mn}}{|\omega_m-\omega_n|^{\epsilon/3}}-\frac{1-\delta_{mn}}{|\omega_n+\omega_m|^{\epsilon/3}} \right)\,.
\ee
The sums can be done in terms of Riemann zeta functions:
\bea
\sum_{m=0}^\infty \frac{1-\delta_{mn}}{|\omega_m-\omega_n|^{\epsilon/3}}&=&\frac{1}{(2\pi T)^{\epsilon/3}}\left[\sum_{m=0}^{n-1}\frac{1}{(n-m)^{\epsilon/3}}+\sum_{m=n+1}^\infty\,\frac{1}{(m-n)^{\epsilon/3}} \right]\nonumber\\
&=&\frac{1}{(2\pi T)^{\epsilon/3}}\left[\sum_{m=0}^{n-1}\frac{1}{(n-m)^{\epsilon/3}}+ \zeta(\epsilon/3)\right]\\
\sum_{m=0}^\infty \frac{1-\delta_{mn}}{|\omega_n+\omega_m|^{\epsilon/3}} &=&\frac{1}{(2\pi T)^{\epsilon/3}}\,\sum_{m=0}^\infty \frac{1}{(m+n+1)^{\epsilon/3}}-\frac{1}{|2\omega_n|^{\epsilon/3}}\nonumber\\
&=&\frac{1}{(2\pi T)^{\epsilon/3}}\,\zeta(\epsilon/3,n+1)-\frac{1}{|2\omega_n|^{\epsilon/3}}\,.
\eea
Shifting variables in the sum, it is easy to check that
\be\label{eq:zetasum}
 \zeta(\epsilon/3)-\zeta(\epsilon/3,n+1)=\sum_{m=1}^n\,\frac{1}{m^{\epsilon/3}}=\sum_{m=0}^{n-1}\frac{1}{(n-m)^{\epsilon/3}}\,.
\ee
We conclude that
\be\label{app:SigmaT}
\Sigma(\omega_n) \approx\,\frac{3\alpha}{\epsilon} \pi T\,\frac{v^\epsilon}{\omega_n^\epsilon}+\frac{3\alpha}{\epsilon}\,(2\pi T)^{1-\epsilon/3}\left[  \zeta(\epsilon/3)-\zeta(\epsilon/3,n+1)\right]\,.
\ee

\subsection{Spectral density of Landau damped bosons}

For a field with propagator $G(\omega, q)$, the spectral density $\rho$ is defined as
\be
G(i\omega_n, q) = \int_{-\infty}^\infty \frac{dq_0}{2\pi}\,\frac{\rho(q_0,q)}{q_0-i \omega_n}
\ee
or
\be
\rho(p_0, p)= 2 \text{Im}[G(p_0+i\delta ,p)]
\ee
as $\delta \to 0^+$. This uses the property $\frac{1}{x+i \delta}= P \frac{1}{x}-i \pi \delta(x)$, with $P$ the principal value. In this section only, we find it convenient to write the euclidean correlator as a function of $i \omega_n$, while in the main text we wrote $G(\omega_n, q)$.

Let us calculate the spectral density for an overdamped boson. The steps are very similar to the analog case in finite density gauge theory \cite{Pisarski:1989cs}. The one loop boson propagator resumming fermion bubbles (and setting the speed $c=1$) is
\be
D(i\omega_n, q) = \frac{1}{\omega_n^2 +q^2 -\frac{M_D^2}{\pi}\frac{i \omega_n}{q}\,\log\frac{i \omega_n-vq}{i \omega_n+vq}}\,.
\ee
In the complex plane, $D(z, q)$
has two simple poles at
\be
z= \pm \omega_T(q)\;,\;\omega_T^2 -q^2 +\frac{M_D^2}{\pi}\frac{\omega_T}{q}\,\log\frac{\omega_T-vq}{\omega_T+vq}=0
\ee
and a branch cut in $z \,\in \,(-vq, \,vq)$.

Given the previous singularity structure, the spectral density is of the form
\be
\rho_B(q_0,q) = \rho_{res}(q) \delta(q_0-\omega_T)+\rho_{cut}(q_0, q) \,\Theta(vq-|q_0|)\,.
\ee
The first term is determined by the residue of $D(z,q)$ at $z=\omega_T$:
\be
\rho_{res}(q)=-2\pi \left[ \frac{\partial D^{-1}}{\partial q_0}\Big|_{q_0=\omega_T}\right]^{-1}=2\pi \frac{\omega_T(\omega_T^2-v^2 q^2)}{(\omega_T^2+q^2)(\omega_T^2-v^2 q^2)+(2/\pi) M_D^2 v \omega_T^2}\,.
\ee
Note that when $q \to 0$, $\omega_T^2 \sim v M_D^2$ and $\rho_{res}\sim \frac{1}{\omega_T}$. In the opposite limit $q\to \infty$, $\omega_T \sim q$ and $\rho_{res} \sim 1/q$. In both cases, the residues occur at very high scales and $\rho_{res}$ is inversely proportional to the high scale. So such poles contribute negligibly to the quantum effects at low energies, and will be ignored.

Let us then focus on the contribution from the cut, which encodes the effects of Landau damping. The discontinuity comes from the branch cut in the logarithm, which gives
\be
\rho_{cut}(q_0, q) = 2\,\frac{M_D^2 q_0/q}{\left(q_0^2-q^2+\frac{M_D^2}{\pi}\,\frac{q_0}{q}\,\log\frac{vq-q_0}{vq+q_0}\right)^2+\left(M_D^2 \frac{q_0}{q}\right)^2}\,.
\ee
In the $z=3$ regime, this becomes
\be
\rho_{cut}(q_0, q) \approx 2\,\frac{M_D^2 q_0/q}{q^4+\left(M_D^2 \frac{q_0}{q}\right)^2}\,.
\ee
Note that
\be
 \int_0^{v q}\,\frac{dq_0}{\pi q_0}\,\rho_{cut}(q_0,q)\approx \frac{1}{q^2}\,,
\ee
for $q \ll M_D$, and hence (\ref{eq:density}) holds.

\subsection{Phase structure and self-consistent $z=3$ scaling}

In this appendix we include some details for deriving the phases in \S \ref{sec:implications} as well as for verifying the validity of the scalar $z=3$ scaling, with fermion self-energy back-reaction at finite T. 

To begin, the fermion self-energy can be approximated by the following ansatz: 
\be\label{eq: sigma_ansatz}
\Sigma(\omega)=\Sigma_T+\Lambda^{2\gamma}_{NFL}\omega^{1-2\gamma},\;\;\;\Sigma_T \approx \begin{cases}
\left(\Lambda^{\epsilon/3}_{NFL} M_D^{2\epsilon/3}\pi T\right)^{\frac{1}{1+\epsilon}},& \pi T \ll \Lambda_{NFL}^{1/3}M_D^{2/3}\\
\Lambda^{\epsilon/3}_{NFL} M_D^{2\epsilon/3} \left(\pi T\right)^{1-\epsilon},& \pi T \gg \Lambda_{NFL}^{1/3}M_D^{2/3}\\
\end{cases}\,,
\ee
where $\gamma=\epsilon/3$. Next we investigate the regime structure of the fermionic kinetic function $\omega Z(\omega)=\omega+\Sigma(\omega)$ as a function of $\omega$. Since $\Lambda_{NFL}\ll M_D$, one possibility is the following:
\begin{eqnarray}\label{eq: self-energy-phase-1}
\omega Z(\omega)=\begin{cases}
\;\omega,   & \Lambda_{NFL}\ll \omega \ll M_D\\
\;\Lambda^{2\gamma}_{NFL}\omega^{1-2\gamma}, &\Lambda^2_T\ll \omega \ll \Lambda_{NFL}\\
\;\Sigma_T, & \pi T \ll \omega \ll \Lambda^2_T\\
\end{cases} \,,
\end{eqnarray} 
where the cross-over scale is given by: $\Lambda^2_T=\left(\Sigma_T\Lambda^{-2\gamma}_{NFL}\right)^{\frac{1}{1-2\gamma}}$. The phase structure (\ref{eq: self-energy-phase-1}) is valid only if the intermediate regime exists: $\Lambda^2_T\ll \Lambda_{NFL} \to \Sigma_{T}\ll \Lambda_{NFL}$. This picks the first branch of $\Sigma_T\approx \left(\Lambda^{\epsilon/3}_{NFL} M_D^{2\epsilon/3}\pi T\right)^{\frac{1}{1+\epsilon}}$ in (\ref{eq: sigma_ansatz}) and further constrains: 
\be 
\pi T \ll \Lambda_{NFL}\left(\frac{\Lambda_{NFL}}{M_D}\right)^{2\epsilon/3}\,,
\ee
which also guarantees the existence of the third regime. In the case of $\pi T \gg \Lambda_{NFL}\left(\frac{\Lambda_{NFL}}{M_D}\right)^{2\epsilon/3}$, the intermediate regime disappear, the next possibility is the following: 
\be\label{eq: self-energy-phase-2}
\omega Z\left(\omega\right) = \begin{cases}
\omega & \Sigma_T \ll \omega \ll M_D\\
\Sigma_T = \left(\Lambda^{\epsilon/3}_{NFL} M_D^{2\epsilon/3}\pi T\right)^{\frac{1}{1+\epsilon}}, & \pi T \ll \omega \ll \Sigma_T\\
\end{cases}
\ee
This is valid if $\pi T \ll \Sigma_T \ll M_D$, which is satisfied if 
\be
\pi T \ll \Lambda^{1/3}_{NFL}M_D^{2/3}
\ee 
For $\pi T \gg \Lambda^{1/3}_{NFL}M_D^{2/3}$, there is only a single regime spanning the whole $\pi T \ll \omega \ll M_D$, which corresponds to the Fermi-liquid phase: 
\be\label{eq: self-energy-phase-3}
\omega Z\left(\omega\right) = \omega
\ee

Next we examine the validity of $z=3$ scaling. The scalar self-energy is given by: 
\be
\Pi(\omega_n, q)= \left(\frac{g^2 k_F^2}{2\pi^2 v}\right)\frac{1}{qv}\frac{2\pi}{\beta}\sum^{|n|-1}_{m=0}\tan^{-1}\left(\frac{qv}{\omega_{|n|}+\Sigma\left(\omega_{|n|}-\omega_m\right)+\Sigma\left(\omega_m\right)}\right)
\ee
As long as the inverse tangent function can be approximated by its limiting value $\pi/2$ for all $0\leq m\leq |n|-1$, the Matsubara sum simply produces a factor of $\omega_n$, and the $z=3$ scaling is consistent. To proceed, we assume $q\sim M_D^{2/3}\omega^{1/3}$ and take $\omega_{|n|}+\Sigma\left(\omega_{|n|}-\omega_m\right)+\Sigma\left(\omega_m\right)\sim \omega_{|n|} Z\left(\omega_{|n|}\right)$ for all $0\leq m\leq |n|-1$, since in all cases $\Sigma\left(\omega\right)$ is a non-decreasing function of $\omega$. Self-consistent $z=3$ scaling is then satisfied if $M_D^{2/3}\omega^{1/3}\gg \omega Z\left(\omega\right)$, assuming that $v\sim \mathcal{O}(1)$. We then check for each temperature regime (\ref{eq: self-energy-phase-1}), (\ref{eq: self-energy-phase-2}) and (\ref{eq: self-energy-phase-3}), whether $z=3$ scaling is self-consistent for the entire range $\pi T \ll \omega \ll M_D$.  
\begin{itemize}
\item $\boldsymbol{\Lambda^{1/3}_{NFL}M_D^{2/3}\ll \pi T\ll M_D}$: in this case $\omega Z\left(\omega\right)\sim \omega$ for the entire range $\pi T \ll \omega \ll M_D$. Self-consistent $z=3$ scaling requires that $\omega\ll M_D$, which always obeyed.

\item $\boldsymbol{\Lambda_{NFL} \left(\frac{\Lambda_{NFL}}{M_D}\right)^{2\epsilon/3}\ll \pi T \ll \Lambda^{1/3}_{NFL}M_D^{2/3}}$: in this case $\omega Z(\omega)\sim \omega$ for $\left(\Lambda^{\epsilon/3}_{NFL}M_D^{2\epsilon/3}\pi T\right)^{\frac{1}{1+\epsilon}}\ll \omega\ll M_D$, in which self-consistent $z=3$ scaling requires that $\omega\ll M_D$; going further down $\omega Z(\omega) \sim \left(\Lambda^{\epsilon/3}_{NFL}M_D^{2\epsilon/3}\pi T\right)^{\frac{1}{1+\epsilon}}$ for $\pi T \ll \omega\ll \left(\Lambda^{\epsilon/3}_{NFL}M_D^{2\epsilon/3}\pi T\right)^{\frac{1}{1+\epsilon}}$, in which self-consistent $z=3$ scaling requires that $\omega \gg \left(\Lambda^\epsilon_{NFL}\frac{(\pi T)^3}{M_D^2}\right)^{\frac{1}{1+\epsilon}}$. Both regions are contained by the respective self-consistent $z=3$ constraints in this temperature regime.

\item $\boldsymbol{\pi T \ll \Lambda_{NFL}\left(\frac{\Lambda_{NFL}}{M_D}\right)^{2\epsilon/3}}$: in this case $\omega Z(\omega)\sim \omega$ for $\Lambda_{NFL}\ll \omega \ll M_D$, with self-consistent $z=3$ scaling requires that $\omega \ll M_D$; going down $\omega Z(\omega)\sim \Lambda^{2\gamma}_{NFL}\omega^{1-2\gamma}$ for $\Lambda^2_T \ll \omega \ll \Lambda_{NFL}$, self-consistent $z=3$ scaling requires that $\omega \ll \left(M_D \Lambda^{-\epsilon}_{NFL}\right)^{\frac{1}{1-\epsilon}}$; going down further $\omega Z(\omega)\sim \left(\Lambda^{\epsilon/3}_{NFL} M_D^{2\epsilon/3}\pi T\right)^{\frac{1}{1+\epsilon}}$ for $\pi T \ll \omega \ll \Lambda^2_T$, self-consistent $z=3$ scaling requires that $\omega \gg \left(\Lambda^\epsilon_{NFL}\frac{(\pi T)^3}{M_D^2}\right)^{\frac{1}{1+\epsilon}}$. Again all three regions are contained by their respective self-consistent $z=3$ constraints in this temperature regime. 
\end{itemize}
In summary, we have checked that for $\pi T\ll M_D$, the fermion self-energy ansatz (\ref{eq: sigma_ansatz}) is consistent with $z=3$ scaling over the entire range $\pi T\ll \omega\ll M_D$. 

\bibliography{NFL}

\begin{thebibliography}{27}%
\makeatletter
\providecommand \@ifxundefined [1]{%
 \@ifx{#1\undefined}
}%
\providecommand \@ifnum [1]{%
 \ifnum #1\expandafter \@firstoftwo
 \else \expandafter \@secondoftwo
 \fi
}%
\providecommand \@ifx [1]{%
 \ifx #1\expandafter \@firstoftwo
 \else \expandafter \@secondoftwo
 \fi
}%
\providecommand \natexlab [1]{#1}%
\providecommand \enquote  [1]{``#1''}%
\providecommand \bibnamefont  [1]{#1}%
\providecommand \bibfnamefont [1]{#1}%
\providecommand \citenamefont [1]{#1}%
\providecommand \href@noop [0]{\@secondoftwo}%
\providecommand \href [0]{\begingroup \@sanitize@url \@href}%
\providecommand \@href[1]{\@@startlink{#1}\@@href}%
\providecommand \@@href[1]{\endgroup#1\@@endlink}%
\providecommand \@sanitize@url [0]{\catcode `\\12\catcode `\$12\catcode
  `\&12\catcode `\#12\catcode `\^12\catcode `\_12\catcode `\%12\relax}%
\providecommand \@@startlink[1]{}%
\providecommand \@@endlink[0]{}%
\providecommand \url  [0]{\begingroup\@sanitize@url \@url }%
\providecommand \@url [1]{\endgroup\@href {#1}{\urlprefix }}%
\providecommand \urlprefix  [0]{URL }%
\providecommand \Eprint [0]{\href }%
\providecommand \doibase [0]{http://dx.doi.org/}%
\providecommand \selectlanguage [0]{\@gobble}%
\providecommand \bibinfo  [0]{\@secondoftwo}%
\providecommand \bibfield  [0]{\@secondoftwo}%
\providecommand \translation [1]{[#1]}%
\providecommand \BibitemOpen [0]{}%
\providecommand \bibitemStop [0]{}%
\providecommand \bibitemNoStop [0]{.\EOS\space}%
\providecommand \EOS [0]{\spacefactor3000\relax}%
\providecommand \BibitemShut  [1]{\csname bibitem#1\endcsname}%
\let\auto@bib@innerbib\@empty
%</preamble>
\bibitem [{\citenamefont {Stewart}(2001)}]{Stewart2001}%
  \BibitemOpen
  \bibfield  {author} {\bibinfo {author} {\bibfnamefont {G.~R.}\ \bibnamefont
  {Stewart}},\ }\bibfield  {title} {\enquote {\bibinfo {title}
  {Non-fermi-liquid behavior in $d$- and $f$-electron metals},}\ }\href
  {\doibase 10.1103/RevModPhys.73.797} {\bibfield  {journal} {\bibinfo
  {journal} {Rev. Mod. Phys.}\ }\textbf {\bibinfo {volume} {73}},\ \bibinfo
  {pages} {797--855} (\bibinfo {year} {2001})}\BibitemShut {NoStop}%
\bibitem [{\citenamefont {Broun}(2008)}]{Broun2008}%
  \BibitemOpen
  \bibfield  {author} {\bibinfo {author} {\bibfnamefont {DM}~\bibnamefont
  {Broun}},\ }\bibfield  {title} {\enquote {\bibinfo {title} {What lies beneath
  the dome?}}\ }\href@noop {} {\bibfield  {journal} {\bibinfo  {journal}
  {Nature Physics}\ }\textbf {\bibinfo {volume} {4}},\ \bibinfo {pages}
  {170--172} (\bibinfo {year} {2008})}\BibitemShut {NoStop}%
\bibitem [{\citenamefont {Gegenwart}\ \emph {et~al.}(2008)\citenamefont
  {Gegenwart}, \citenamefont {Si},\ and\ \citenamefont
  {Steglich}}]{Gegenwart2008}%
  \BibitemOpen
  \bibfield  {author} {\bibinfo {author} {\bibfnamefont {P.}~\bibnamefont
  {Gegenwart}}, \bibinfo {author} {\bibfnamefont {Q.}~\bibnamefont {Si}}, \
  and\ \bibinfo {author} {\bibfnamefont {F.}~\bibnamefont {Steglich}},\
  }\bibfield  {title} {\enquote {\bibinfo {title} {Quantum criticality in
  heavy-fermion metals},}\ }\href@noop {} {\bibfield  {journal} {\bibinfo
  {journal} {nature physics}\ }\textbf {\bibinfo {volume} {4}},\ \bibinfo
  {pages} {186--197} (\bibinfo {year} {2008})}\BibitemShut {NoStop}%
\bibitem [{\citenamefont {{Taillefer}}(2010)}]{taillefer-review}%
  \BibitemOpen
  \bibfield  {author} {\bibinfo {author} {\bibfnamefont {L.}~\bibnamefont
  {{Taillefer}}},\ }\bibfield  {title} {\enquote {\bibinfo {title} {{Scattering
  and Pairing in Cuprate Superconductors}},}\ }\href {\doibase
  10.1146/annurev-conmatphys-070909-104117} {\bibfield  {journal} {\bibinfo
  {journal} {Annual Review of Condensed Matter Physics}\ }\textbf {\bibinfo
  {volume} {1}},\ \bibinfo {pages} {51--70} (\bibinfo {year} {2010})},\ \Eprint
  {http://arxiv.org/abs/1003.2972} {arXiv:1003.2972 [cond-mat.supr-con]}
  \BibitemShut {NoStop}%
\bibitem [{\citenamefont {Shibauchi}\ \emph {et~al.}(2013)\citenamefont
  {Shibauchi}, \citenamefont {Carrington},\ and\ \citenamefont
  {Matsuda}}]{Shibauchi2013}%
  \BibitemOpen
  \bibfield  {author} {\bibinfo {author} {\bibfnamefont {T.}~\bibnamefont
  {Shibauchi}}, \bibinfo {author} {\bibfnamefont {A.}~\bibnamefont
  {Carrington}}, \ and\ \bibinfo {author} {\bibfnamefont {Y.}~\bibnamefont
  {Matsuda}},\ }\bibfield  {title} {\enquote {\bibinfo {title} {Quantum
  critical point lying beneath the superconducting dome in iron-pnictides},}\
  }\href@noop {} {\bibfield  {journal} {\bibinfo  {journal} {arXiv preprint
  arXiv:1304.6387}\ } (\bibinfo {year} {2013})}\BibitemShut {NoStop}%
\bibitem [{\citenamefont {Metlitski}\ \emph {et~al.}(2015)\citenamefont
  {Metlitski}, \citenamefont {Mross}, \citenamefont {Sachdev},\ and\
  \citenamefont {Senthil}}]{Metlitski}%
  \BibitemOpen
  \bibfield  {author} {\bibinfo {author} {\bibfnamefont {M.}~\bibnamefont
  {Metlitski}}, \bibinfo {author} {\bibfnamefont {D.}~\bibnamefont {Mross}},
  \bibinfo {author} {\bibfnamefont {S.}~\bibnamefont {Sachdev}}, \ and\
  \bibinfo {author} {\bibfnamefont {T.}~\bibnamefont {Senthil}},\ }\bibfield
  {title} {\enquote {\bibinfo {title} {Cooper pairing in non-fermi liquids},}\
  }\href {\doibase 10.1103/PhysRevB.91.115111} {\bibfield  {journal} {\bibinfo
  {journal} {Phys. Rev. B}\ }\textbf {\bibinfo {volume} {91}},\ \bibinfo
  {pages} {115111} (\bibinfo {year} {2015})}\BibitemShut {NoStop}%
\bibitem [{\citenamefont {Fitzpatrick}\ \emph {et~al.}(2013)\citenamefont
  {Fitzpatrick}, \citenamefont {Kachru}, \citenamefont {Kaplan},\ and\
  \citenamefont {Raghu}}]{Fitzpatrickone}%
  \BibitemOpen
  \bibfield  {author} {\bibinfo {author} {\bibfnamefont {A.}~\bibnamefont
  {Fitzpatrick}}, \bibinfo {author} {\bibfnamefont {S.}~\bibnamefont {Kachru}},
  \bibinfo {author} {\bibfnamefont {J.}~\bibnamefont {Kaplan}}, \ and\ \bibinfo
  {author} {\bibfnamefont {S.}~\bibnamefont {Raghu}},\ }\bibfield  {title}
  {\enquote {\bibinfo {title} {Non-fermi-liquid fixed point in a wilsonian
  theory of quantum critical metals},}\ }\href {\doibase
  10.1103/PhysRevB.88.125116} {\bibfield  {journal} {\bibinfo  {journal} {Phys.
  Rev. B}\ }\textbf {\bibinfo {volume} {88}},\ \bibinfo {pages} {125116}
  (\bibinfo {year} {2013})}\BibitemShut {NoStop}%
\bibitem [{\citenamefont {{Fitzpatrick}}\ \emph {et~al.}(2014)\citenamefont
  {{Fitzpatrick}}, \citenamefont {{Kachru}}, \citenamefont {{Kaplan}},\ and\
  \citenamefont {{Raghu}}}]{FKKRtwo}%
  \BibitemOpen
  \bibfield  {author} {\bibinfo {author} {\bibfnamefont {A.~L.}\ \bibnamefont
  {{Fitzpatrick}}}, \bibinfo {author} {\bibfnamefont {S.}~\bibnamefont
  {{Kachru}}}, \bibinfo {author} {\bibfnamefont {J.}~\bibnamefont {{Kaplan}}},
  \ and\ \bibinfo {author} {\bibfnamefont {S.}~\bibnamefont {{Raghu}}},\
  }\bibfield  {title} {\enquote {\bibinfo {title} {{Non-Fermi-liquid behavior
  of large-N$_{B}$ quantum critical metals}},}\ }\href {\doibase
  10.1103/PhysRevB.89.165114} {\bibfield  {journal} {\bibinfo  {journal}
  {\prb}\ }\textbf {\bibinfo {volume} {89}},\ \bibinfo {eid} {165114} (\bibinfo
  {year} {2014})},\ \Eprint {http://arxiv.org/abs/1312.3321} {arXiv:1312.3321
  [cond-mat.str-el]} \BibitemShut {NoStop}%
\bibitem [{\citenamefont {Torroba}\ and\ \citenamefont
  {Wang}(2014)}]{Torroba:2014gqa}%
  \BibitemOpen
  \bibfield  {author} {\bibinfo {author} {\bibfnamefont {G.}~\bibnamefont
  {Torroba}}\ and\ \bibinfo {author} {\bibfnamefont {H.}~\bibnamefont {Wang}},\
  }\bibfield  {title} {\enquote {\bibinfo {title} {{Quantum critical metals in
  $4-\epsilon$ dimensions}},}\ }\href {\doibase 10.1103/PhysRevB.90.165144}
  {\bibfield  {journal} {\bibinfo  {journal} {Phys.Rev.}\ }\textbf {\bibinfo
  {volume} {B90}},\ \bibinfo {pages} {165144} (\bibinfo {year} {2014})},\
  \Eprint {http://arxiv.org/abs/1406.3029} {arXiv:1406.3029 [cond-mat.str-el]}
  \BibitemShut {NoStop}%
%%CITATION = ARXIV:1406.3029;%%
\bibitem [{\citenamefont {Raghu}\ \emph {et~al.}(2015)\citenamefont {Raghu},
  \citenamefont {Torroba},\ and\ \citenamefont {Wang}}]{Raghu:2015sna}%
  \BibitemOpen
  \bibfield  {author} {\bibinfo {author} {\bibfnamefont {S.}~\bibnamefont
  {Raghu}}, \bibinfo {author} {\bibfnamefont {G.}~\bibnamefont {Torroba}}, \
  and\ \bibinfo {author} {\bibfnamefont {H.}~\bibnamefont {Wang}},\ }\bibfield
  {title} {\enquote {\bibinfo {title} {{Metallic quantum critical points with
  finite BCS couplings}},}\ }\href {\doibase 10.1103/PhysRevB.92.205104}
  {\bibfield  {journal} {\bibinfo  {journal} {Phys. Rev.}\ }\textbf {\bibinfo
  {volume} {B92}},\ \bibinfo {pages} {205104} (\bibinfo {year} {2015})},\
  \Eprint {http://arxiv.org/abs/1507.06652} {arXiv:1507.06652
  [cond-mat.str-el]} \BibitemShut {NoStop}%
%%CITATION = ARXIV:1507.06652;%%
\bibitem [{\citenamefont {Wang}\ \emph {et~al.}(2017)\citenamefont {Wang},
  \citenamefont {Raghu},\ and\ \citenamefont {Torroba}}]{Wang:2016hir}%
  \BibitemOpen
  \bibfield  {author} {\bibinfo {author} {\bibfnamefont {H.}~\bibnamefont
  {Wang}}, \bibinfo {author} {\bibfnamefont {S.}~\bibnamefont {Raghu}}, \ and\
  \bibinfo {author} {\bibfnamefont {G.}~\bibnamefont {Torroba}},\ }\bibfield
  {title} {\enquote {\bibinfo {title} {{Non-Fermi liquid Superconductivity:
  Eliashberg versus the Renormalization Group}},}\ }\href {\doibase
  10.1103/PhysRevB.95.165137} {\bibfield  {journal} {\bibinfo  {journal} {Phys.
  Rev.}\ }\textbf {\bibinfo {volume} {B95}},\ \bibinfo {pages} {165137}
  (\bibinfo {year} {2017})},\ \Eprint {http://arxiv.org/abs/1612.01971}
  {arXiv:1612.01971 [cond-mat.str-el]} \BibitemShut {NoStop}%
%%CITATION = ARXIV:1612.01971;%%
\bibitem [{\citenamefont {Torroba}\ \emph {et~al.}()\citenamefont {Torroba},
  \citenamefont {Wang},\ and\ \citenamefont {Wang}}]{future}%
  \BibitemOpen
  \bibfield  {author} {\bibinfo {author} {\bibfnamefont {G.}~\bibnamefont
  {Torroba}}, \bibinfo {author} {\bibfnamefont {H.}~\bibnamefont {Wang}}, \
  and\ \bibinfo {author} {\bibfnamefont {Y.}~\bibnamefont {Wang}},\ }\bibfield
  {title} {\enquote {\bibinfo {title} {{to appear}},}\ }\href@noop {} {\
  }\BibitemShut {NoStop}%
\bibitem [{\citenamefont {Le~Bellac}(2000)}]{le2000thermal}%
  \BibitemOpen
  \bibfield  {author} {\bibinfo {author} {\bibfnamefont {M.}~\bibnamefont
  {Le~Bellac}},\ }\href@noop {} {\emph {\bibinfo {title} {Thermal field
  theory}}}\ (\bibinfo  {publisher} {Cambridge University Press},\ \bibinfo
  {year} {2000})\BibitemShut {NoStop}%
\bibitem [{\citenamefont {Blaizot}\ and\ \citenamefont
  {Iancu}(1997)}]{Blaizot:1996az}%
  \BibitemOpen
  \bibfield  {author} {\bibinfo {author} {\bibfnamefont {Jean-Paul}\
  \bibnamefont {Blaizot}}\ and\ \bibinfo {author} {\bibfnamefont {Edmond}\
  \bibnamefont {Iancu}},\ }\bibfield  {title} {\enquote {\bibinfo {title}
  {{Lifetimes of quasiparticles and collective excitations in hot QED
  plasmas}},}\ }\href {\doibase 10.1103/PhysRevD.55.973} {\bibfield  {journal}
  {\bibinfo  {journal} {Phys. Rev.}\ }\textbf {\bibinfo {volume} {D55}},\
  \bibinfo {pages} {973--996} (\bibinfo {year} {1997})},\ \Eprint
  {http://arxiv.org/abs/hep-ph/9607303} {arXiv:hep-ph/9607303 [hep-ph]}
  \BibitemShut {NoStop}%
%%CITATION = HEP-PH/9607303;%%
\bibitem [{\citenamefont {Abanov}\ \emph {et~al.}(2003)\citenamefont {Abanov},
  \citenamefont {Chubukov},\ and\ \citenamefont
  {Schmalian}}]{abanov2003quantum}%
  \BibitemOpen
  \bibfield  {author} {\bibinfo {author} {\bibfnamefont {A.}~\bibnamefont
  {Abanov}}, \bibinfo {author} {\bibfnamefont {A.~V.}\ \bibnamefont
  {Chubukov}}, \ and\ \bibinfo {author} {\bibfnamefont {J.}~\bibnamefont
  {Schmalian}},\ }\bibfield  {title} {\enquote {\bibinfo {title}
  {Quantum-critical theory of the spin-fermion model and its application to
  cuprates: normal state analysis},}\ }\href@noop {} {\bibfield  {journal}
  {\bibinfo  {journal} {Advances in Physics}\ }\textbf {\bibinfo {volume}
  {52}},\ \bibinfo {pages} {119--218} (\bibinfo {year} {2003})}\BibitemShut
  {NoStop}%
\bibitem [{\citenamefont {Polchinski}(1994)}]{Polchinski1994}%
  \BibitemOpen
  \bibfield  {author} {\bibinfo {author} {\bibfnamefont {J.}~\bibnamefont
  {Polchinski}},\ }\bibfield  {title} {\enquote {\bibinfo {title} {Low-energy
  dynamics of the spinon-gauge system},}\ }\href@noop {} {\bibfield  {journal}
  {\bibinfo  {journal} {Nuclear Physics B}\ }\textbf {\bibinfo {volume}
  {422}},\ \bibinfo {pages} {617--633} (\bibinfo {year} {1994})}\BibitemShut
  {NoStop}%
\bibitem [{\citenamefont {Altshuler}\ \emph {et~al.}(1994)\citenamefont
  {Altshuler}, \citenamefont {Ioffe},\ and\ \citenamefont
  {Millis}}]{Altshuler1994}%
  \BibitemOpen
  \bibfield  {author} {\bibinfo {author} {\bibfnamefont {B.~L.}\ \bibnamefont
  {Altshuler}}, \bibinfo {author} {\bibfnamefont {L.~B.}\ \bibnamefont
  {Ioffe}}, \ and\ \bibinfo {author} {\bibfnamefont {A.~J.}\ \bibnamefont
  {Millis}},\ }\bibfield  {title} {\enquote {\bibinfo {title} {Low-energy
  properties of fermions with singular interactions},}\ }\href {\doibase
  10.1103/PhysRevB.50.14048} {\bibfield  {journal} {\bibinfo  {journal} {Phys.
  Rev. B}\ }\textbf {\bibinfo {volume} {50}},\ \bibinfo {pages} {14048--14064}
  (\bibinfo {year} {1994})}\BibitemShut {NoStop}%
\bibitem [{\citenamefont {Lee}(2009)}]{Lee2009}%
  \BibitemOpen
  \bibfield  {author} {\bibinfo {author} {\bibfnamefont {S-S.}\ \bibnamefont
  {Lee}},\ }\bibfield  {title} {\enquote {\bibinfo {title} {Low-energy
  effective theory of fermi surface coupled with u(1) gauge field in $2+1$
  dimensions},}\ }\href {\doibase 10.1103/PhysRevB.80.165102} {\bibfield
  {journal} {\bibinfo  {journal} {Phys. Rev. B}\ }\textbf {\bibinfo {volume}
  {80}},\ \bibinfo {pages} {165102} (\bibinfo {year} {2009})}\BibitemShut
  {NoStop}%
\bibitem [{\citenamefont {Shankar}(1994)}]{Shankar}%
  \BibitemOpen
  \bibfield  {author} {\bibinfo {author} {\bibfnamefont {R.}~\bibnamefont
  {Shankar}},\ }\bibfield  {title} {\enquote {\bibinfo {title}
  {{Renormalization group approach to interacting fermions}},}\ }\href
  {\doibase 10.1103/RevModPhys.66.129} {\bibfield  {journal} {\bibinfo
  {journal} {Rev.Mod.Phys.}\ }\textbf {\bibinfo {volume} {66}},\ \bibinfo
  {pages} {129--192} (\bibinfo {year} {1994})}\BibitemShut {NoStop}%
%%CITATION = RMPHA,66,129;%%
\bibitem [{\citenamefont {Nagaosa}(2013)}]{nagaosa2013quantum}%
  \BibitemOpen
  \bibfield  {author} {\bibinfo {author} {\bibfnamefont {N.}~\bibnamefont
  {Nagaosa}},\ }\href@noop {} {\emph {\bibinfo {title} {Quantum field theory in
  condensed matter physics}}}\ (\bibinfo  {publisher} {Springer Science \&
  Business Media},\ \bibinfo {year} {2013})\BibitemShut {NoStop}%
\bibitem [{\citenamefont {Moon}\ and\ \citenamefont
  {Chubukov}(2010)}]{Moon2010}%
  \BibitemOpen
  \bibfield  {author} {\bibinfo {author} {\bibfnamefont {E-G.}\ \bibnamefont
  {Moon}}\ and\ \bibinfo {author} {\bibfnamefont {A.}~\bibnamefont
  {Chubukov}},\ }\bibfield  {title} {\enquote {\bibinfo {title}
  {Quantum-critical pairing with varying exponents},}\ }\href@noop {}
  {\bibfield  {journal} {\bibinfo  {journal} {Journal of Low Temperature
  Physics}\ }\textbf {\bibinfo {volume} {161}},\ \bibinfo {pages} {263--281}
  (\bibinfo {year} {2010})}\BibitemShut {NoStop}%
\bibitem [{Note1()}]{Note1}%
  \BibitemOpen
  \bibinfo {note} {More precisely, there is a subleading $\protect \qopname
  \relax o{log}T$ dependence from expanding (\ref {eq:SigmaT}) near
  $d=2$.}\BibitemShut {Stop}%
\bibitem [{\citenamefont {Meszena}\ \emph {et~al.}(2016)\citenamefont
  {Meszena}, \citenamefont {Saterskog}, \citenamefont {Bagrov},\ and\
  \citenamefont {Schalm}}]{meszena2016nonperturbative}%
  \BibitemOpen
  \bibfield  {author} {\bibinfo {author} {\bibfnamefont {B.}~\bibnamefont
  {Meszena}}, \bibinfo {author} {\bibfnamefont {P.}~\bibnamefont {Saterskog}},
  \bibinfo {author} {\bibfnamefont {A.}~\bibnamefont {Bagrov}}, \ and\ \bibinfo
  {author} {\bibfnamefont {K.}~\bibnamefont {Schalm}},\ }\bibfield  {title}
  {\enquote {\bibinfo {title} {Nonperturbative emergence of non-fermi-liquid
  behavior in d= 2 quantum critical metals},}\ }\href@noop {} {\bibfield
  {journal} {\bibinfo  {journal} {Physical Review B}\ }\textbf {\bibinfo
  {volume} {94}},\ \bibinfo {pages} {115134} (\bibinfo {year}
  {2016})}\BibitemShut {NoStop}%
\bibitem [{\citenamefont {Saterskog}\ \emph {et~al.}(2016)\citenamefont
  {Saterskog}, \citenamefont {Meszena},\ and\ \citenamefont
  {Schalm}}]{saterskog2016exact}%
  \BibitemOpen
  \bibfield  {author} {\bibinfo {author} {\bibfnamefont {P.}~\bibnamefont
  {Saterskog}}, \bibinfo {author} {\bibfnamefont {B.}~\bibnamefont {Meszena}},
  \ and\ \bibinfo {author} {\bibfnamefont {K.}~\bibnamefont {Schalm}},\
  }\bibfield  {title} {\enquote {\bibinfo {title} {The exact spectrum of a $ d=
  2$ quantum critical metal in the limit $ k_f \to \infty$, $ n_f \to 0$ with $
  n_f k _f $ fixed},}\ }\href@noop {} {\bibfield  {journal} {\bibinfo
  {journal} {arXiv preprint arXiv:1612.05326}\ } (\bibinfo {year}
  {2016})}\BibitemShut {NoStop}%
\bibitem [{\citenamefont {Lee}(2017)}]{lee2017recent}%
  \BibitemOpen
  \bibfield  {author} {\bibinfo {author} {\bibfnamefont {S.-S.}\ \bibnamefont
  {Lee}},\ }\bibfield  {title} {\enquote {\bibinfo {title} {Recent developments
  in non-fermi liquid theory},}\ }\href@noop {} {\bibfield  {journal} {\bibinfo
   {journal} {arXiv preprint arXiv:1703.08172}\ } (\bibinfo {year}
  {2017})}\BibitemShut {NoStop}%
\bibitem [{\citenamefont {{Wang}}\ \emph {et~al.}(2016)\citenamefont {{Wang}},
  \citenamefont {{Abanov}}, \citenamefont {{Altshuler}}, \citenamefont
  {{Yuzbashyan}},\ and\ \citenamefont {{Chubukov}}}]{2016PhRvL.117o7001W}%
  \BibitemOpen
  \bibfield  {author} {\bibinfo {author} {\bibfnamefont {Y.}~\bibnamefont
  {{Wang}}}, \bibinfo {author} {\bibfnamefont {A.}~\bibnamefont {{Abanov}}},
  \bibinfo {author} {\bibfnamefont {B.~L.}\ \bibnamefont {{Altshuler}}},
  \bibinfo {author} {\bibfnamefont {E.~A.}\ \bibnamefont {{Yuzbashyan}}}, \
  and\ \bibinfo {author} {\bibfnamefont {A.~V.}\ \bibnamefont {{Chubukov}}},\
  }\bibfield  {title} {\enquote {\bibinfo {title} {{Superconductivity near a
  Quantum-Critical Point: The Special Role of the First Matsubara
  Frequency}},}\ }\href {\doibase 10.1103/PhysRevLett.117.157001} {\bibfield
  {journal} {\bibinfo  {journal} {Physical Review Letters}\ }\textbf {\bibinfo
  {volume} {117}},\ \bibinfo {eid} {157001} (\bibinfo {year} {2016})},\ \Eprint
  {http://arxiv.org/abs/1606.01252} {arXiv:1606.01252 [cond-mat.supr-con]}
  \BibitemShut {NoStop}%
\bibitem [{\citenamefont {Pisarski}(1989)}]{Pisarski:1989cs}%
  \BibitemOpen
  \bibfield  {author} {\bibinfo {author} {\bibfnamefont {R.~D.}\ \bibnamefont
  {Pisarski}},\ }\bibfield  {title} {\enquote {\bibinfo {title} {{Renormalized
  Gauge Propagator in Hot Gauge Theories}},}\ }\bibfield  {booktitle} {\emph
  {\bibinfo {booktitle} {{Physica A158 (1989) 146-157}}},\ }\href@noop {}
  {\bibfield  {journal} {\bibinfo  {journal} {Physica}\ }\textbf {\bibinfo
  {volume} {A158}},\ \bibinfo {pages} {146--157} (\bibinfo {year}
  {1989})}\BibitemShut {NoStop}%
%%CITATION = PHYSA,A158,146;%%
\end{thebibliography}%

\end{document}